\documentclass{iopart}
\usepackage{iopams}
\usepackage{setstack}
\usepackage{color}
\usepackage{graphicx}

\begin{document}

\title{Langevin spin dynamics based on ab initio calculations: numerical schemes and applications}

\author{L R\'ozsa$^1$, L Udvardi$^{1,2}$ and L Szunyogh$^{1,2}$}

\address{$^1$ Department of Theoretical Physics, Budapest University of Technology and Economics, Budafoki \'ut 8,~H-1111 Budapest, Hungary}
\address{$^2$ Condensed Matter Research Group of Hungarian Academy of Sciences, Budapest University of Technology and Economics, Budafoki \'ut 8,~H-1111 Budapest, Hungary }

\ead{rozsa@phy.bme.hu}

\begin{abstract}

A method is proposed to study the finite-temperature behaviour of small magnetic clusters based on solving the stochastic Landau-Lifshitz-Gilbert equations, where the effective magnetic field is calculated directly during the solution of the dynamical equations from first principles instead of relying on an effective spin Hamiltonian. Different numerical solvers are discussed in the case of a one-dimensional Heisenberg chain with nearest-neighbour interactions. We performed detailed investigations for a monatomic chain of ten Co atoms on top of Au($001$) surface. We found a spiral-like ground state of the spins due to Dzyaloshinsky-Moriya interactions, while the finite-temperature magnetic behaviour of the system was well described by a nearest-neighbour Heisenberg model including easy-axis anisotropy.

\end{abstract}


\submitto{\JPCM}

\maketitle

\section{Introduction}

The study of low-dimensional magnetic systems is at the center of current research interest because of their applicability in memory and spintronics devices. Various experimental techniques, such as spin-polarized scanning tunneling microscopy\cite{Wiesendanger}, have made it possible to determine the magnetic structure of systems down to the atomic level. Magnetic devices can often be successfully modelled by continuum micromagnetic methods\cite{Aharoni, Kronmuller}.
Atomistic spin dynamics simulations provide a way to theoretically model magnetic systems containing from several atoms to a few thousand atoms, on time scales ranging from a few femtoseconds to several hundred picoseconds\cite{Nowak}.
Most of these methods are based on the numerical solution of the  stochastic Landau-Lifshitz-Gilbert (LLG) equation\cite{Landau,Gilbert,Brown,Kubo}, where the torque acting on the spin vectors is determined from a
generalized Heisenberg model with parameters obtained from ab initio calculations\cite{Skubic,Liechtenstein,Udvardi}.

While in case of bulk systems or thin films with at least tetragonal symmetry the construction of the effective Hamiltonian is straightforward\cite{Udvardi}, in small magnetic clusters the reduced symmetry of the system makes this task quite complicated. This concerns, in particular, the on-site magnetic anisotropy and the off-diagonal matrix elements of the exchange tensor. These terms of the effective Hamiltonian are related to the relativistic spin-orbit coupling, therefore their role is essential in spintronics applications. In order to avoid this technical problem of ab initio based spin models, first principles spin dynamics has to be used, where the effective field driving the motion of the spins is calculated directly from density functional theory.

The foundation of first principles spin dynamics in itinerant-electron systems was laid down by Antropov \etal\cite{Antropov1995,Antropov1996} and was later developed to include Berry phase effects\cite{Niu1999} and many-body effects in terms of time-dependent spin-density functional theory\cite{Capelle2001}. It was pointed out that the adiabatic decoupling of the motion of the magnetization averaged over an atomic volume and the electronic degrees of freedom results in an equation identical to the Landau-Lifshitz-Gilbert equation. The time evolution of the atomic magnetization can be treated similarly to the description of the motion of the nuclei in molecular dynamics. In molecular dynamics the forces are calculated by means of ab initio methods but the classical equation of motion is solved. In spin dynamics the torque driving the motion of the atomic moments is calculated from first principles and it is used to determine the orientation of the magnetization at the next time step via the classical Landau-Lifshitz-Gilbert equation.

One realization of ab initio spin dynamics is based on the constrained local moment (CLM) approach proposed by Stocks \etal\cite{Stocks1998, Stocks1999} following the constrained density functional theory developed by Dederichs \etal\cite{Dederichs}. In the constrained local moment method the Kohn-Sham equations are solved in the presence of a constraining field ensuring that the local moments point to predefined directions. The opposite of this constraining field is the internal effective field which rotates the spins, therefore it should be used in the Landau-Lifshitz-Gilbert equations.

In the present work the effective field is determined relying on the magnetic force theorem\cite{Liechtenstein,Jansen}. By using multiple scattering theory, analytic formulas are derived for the derivatives of the band energy with respect to the transverse change of the exchange field. The electronic structure of the system is determined by applying the embedded cluster method in the framework of the fully relativistic Korringa-Kohn-Rostoker method\cite{Lazarovits}.
Since the Landau-Lifshitz-Gilbert equations are rewritten into a form appropriate for our ab initio calculations, a new numerical method was implemented, based on the one proposed by Mentink \etal\cite{Mentink}.

The new numerical scheme is first tested on a model Hamiltonian describing a linear chain of atoms with ferromagnetic nearest-neighbour Heisenberg coupling. The model was chosen since it has an analytic solution\cite{Shubin,Fisher}, therefore the numerical results can be compared to exact values. Another reason for studying this model is that linear chains of atoms are of great interest. Special non-collinear ground states were reported experimentally for Fe/Ir($001$)\cite{Menzel} as well as theoretically for Mn/Ni($001$)\cite{Lounis}.  The magnetism of monatomic Co chains on a Pt($997$) surface has been studied in detail in \cite{Gambardella,Dallmeyer}. Ab initio calculations were performed for free-standing infinite Co chains\cite{Tung2007,Nautiyal} as well as for those supported by Pt or Cu surfaces\cite{Tung2011} or embedded in carbon
nanotubes\cite{Xie}. It was found by Hong \etal\cite{Hong} that, although the system is always
ferromagnetic, the anisotropy prefers the chain direction in the supported Co/Cu($001$) case and the perpendicular direction in the free-standing case. It was shown by Tung \etal\cite{Tung2011B} and later by T\"ows \etal\cite{Tows} that this system does not have a spin spiral ground state, contrary to V, Mn and Fe chains, where the spiral ordering is the consequence of frustrated exchange interactions. Finite chains have also been studied by ab initio calculations\cite{Vindigni,Guirado,Lazarovits2003,Ujfalussy2004}.

In section \ref{efff} the calculation of the effective field appearing in the Landau-Lifshitz-Gilbert equation is detailed. In section \ref{numint} three numerical integration schemes are described for solving the dynamical equations in the local coordinate system. Based on model calculations described in section \ref{model}, it is concluded that the so-called one-step scheme has the most advantageous properties out of the three integration schemes. In section \ref{Co10Au001} the ab initio method is applied to a linear chain of ten Co atoms deposited on Au($001$) and it is compared to a model Hamiltonian containing Heisenberg exchange interactions and uniaxial magnetic anisotropy. It is found that the system is ferromagnetic and the magnetic anisotropy prefers the chain direction, in agreement with earlier calculations carried out for Cu($001$) surface\cite{Hong,Tung2011}. On the other hand, due to the Dzyaloshinsky-Moriya interactions\cite{Dzyaloshinsky,Moriya}
the ground state of the system turned out to resemble a spin spiral state. It was found that the temperature-dependent energy and magnetization curves are well described by a nearest-neighbour Heisenberg model, while the simulated switching time between the degenerate ground states can also be satisfactorily reproduced in terms of the simple spin model containing additional on-site anisotropy terms.

\section{Calculating the effective field in the stochastic Landau-Lifshitz-Gilbert equation\label{efff}}

In case of atomistic simulations, the stochastic Landau-Lifshitz-Gilbert equation has the form
\begin{eqnarray}
\frac{\partial \boldsymbol{M}_{i}}{\partial t}&=&-\gamma'\boldsymbol{M}_{i}\times(\boldsymbol{B}^{eff}_{i}+\boldsymbol{B}_{i}^{th} )\nonumber
\\
&& - \frac{\alpha\gamma'}{M_{i}}\boldsymbol{M}_{i}\times \Big[\boldsymbol{M}_{i}\times(\boldsymbol{B}^{eff}_{i}+\boldsymbol{B}_{i}^{th} )\Big],\label{LLG}
\\
\boldsymbol{B}^{eff}_{i}&=&-\frac{\partial E}{\partial \boldsymbol{M}_{i}}=-\frac{1}{M_{i}} \frac{\partial E}{\partial \boldsymbol{\sigma}_{i}},
\\
\boldsymbol{B}_{i}^{th} &=&\sqrt{2D_{i}} \circ \boldsymbol{\eta}_{i}=\sqrt{\frac{2\alpha k_{B}T}{M_{i}\gamma}} \circ \boldsymbol{\eta}_{i},\label{deviation}
\end{eqnarray}
where $\boldsymbol{M}_{i}=M_{i}\boldsymbol{\sigma}_{i}$ stands for the localized magnetic moment (spin) at site $i$, 
$\alpha$ is the Gilbert damping, $\gamma'=\frac{\gamma}{1+\alpha^{2}}$ with the gyromagnetic factor  $\gamma=\frac{2 \mu_B}{\hbar}=\frac{e}{m}$. For the stochastic part (\ref{deviation}), $T$ denotes the temperature and $\boldsymbol{\eta}_{i}$ is the white noise. The $\circ$ symbol denotes that the Stratonovich interpretation of the stochastic differential equation was used, which is necessary to preserve the magnitude $M_{i}$ of the spin during the time evolution\cite{Mentink}, as well as to satisfy the correct thermal equilibrium distribution for the spins\cite{Palacios}.
This quasiclassical approach may provide a suitable description of the time evolution of the spins if the electronic processes are considerably faster than the motion of the localized moments\cite{Halilov}.

\begin{figure}
\centering
\includegraphics[width=\columnwidth]{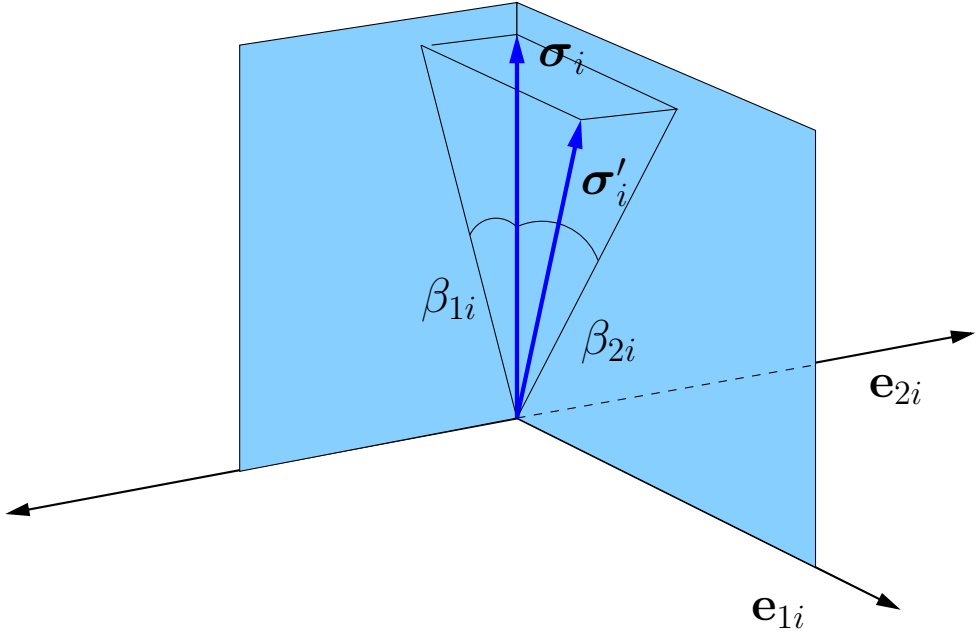}
\caption{Sketch of the spin vector $\boldsymbol{\sigma}_{i}$, the unit vectors $\boldsymbol{e}_{1i}, \boldsymbol{e}_{2i}$ and the angle variables $\beta_{1i}, \beta_{2i}$ as introduced in the text. The vector $\boldsymbol{\sigma}^\prime_{i}$ represents the spin after an infinitesimal rotation.}
\label{angles}
\end{figure}

By using the energy of the system $E$ from ab initio calculations, the effective field $\boldsymbol{B}_{i}^{eff}$ is determined  in the local coordinate system, which transforms along with the spin vectors $\boldsymbol{\sigma}_{i}$. Introducing the unit vectors $\boldsymbol{e}_{1i}, \boldsymbol{e}_{2i}$, as well as the angles describing the infinitesimal rotations around these vectors $\beta_{1i}, \beta_{2i}$ as in figure \ref{angles} and making use of the identities
\begin{eqnarray}
\boldsymbol{\sigma}_{i}\boldsymbol{e}_{1i}=\boldsymbol{\sigma}_{i}\boldsymbol{e}_{2i}=\boldsymbol{e}_{1i}\boldsymbol{e}_{2i}=0,
\\
\boldsymbol{e}_{1i}\times\boldsymbol{\sigma}_{i}=-\boldsymbol{e}_{2i}, \qquad \boldsymbol{e}_{2i}\times\boldsymbol{\sigma}_{i}=\boldsymbol{e}_{1i}, \label{e1e2}
\\
\rmd\boldsymbol{\sigma}_{i}=-\rmd\beta_{1i}\boldsymbol{e}_{2i} + \rmd\beta_{2i}\boldsymbol{e}_{1i}, \label{dsigma}
\\
\boldsymbol{B}^{eff}_{i\bot}=\frac{1}{M_{i}} \frac{\partial E}{\partial \beta_{1i}}\boldsymbol{e}_{2i} - \frac{1}{M_{i}} \frac{\partial E}{\partial \beta_{2i}}\boldsymbol{e}_{1i},
\end{eqnarray}
the stochastic Landau-Lifshitz-Gilbert equations transform into
\begin{eqnarray}
\rmd\beta_{2i}&=&\frac{\gamma'}{M_{i}}\frac{\partial E}{\partial \beta_{1i}}\rmd t - \alpha\frac{\gamma'}{M_{i}}\frac{\partial E}{\partial \beta_{2i}}\rmd t \nonumber
\\
&&+ \gamma' \sqrt{2D_{i}}\boldsymbol{e}_{2i}\circ \rmd\boldsymbol{W}_{i} + \alpha\gamma' \sqrt{2D_{i}}\boldsymbol{e}_{1i}\circ \rmd\boldsymbol{W}_{i}, \label{LLGloc1}
\\
\rmd\beta_{1i}&=&-\frac{\gamma'}{M_{i}}\frac{\partial E}{\partial \beta_{2i}}\rmd t - \alpha\frac{\gamma'}{M_{i}}\frac{\partial E}{\partial \beta_{1i}}\rmd t \nonumber
\\
&&+ \gamma' \sqrt{2D_{i}}\boldsymbol{e}_{1i}\circ \rmd\boldsymbol{W}_{i} - \alpha\gamma' \sqrt{2D_{i}}\boldsymbol{e}_{2i}\circ \rmd\boldsymbol{W}_{i}, \label{LLGloc2}
\end{eqnarray}
with $\rmd\boldsymbol{W}_{i}$ the infinitesimal form of the Wiener process with the usual properties\cite{Kloeden}: an almost surely continuous Gaussian stochastic process starting from $W_{i}^{r}(0)=0$ with first and second moments $\langle W_{i}^{r}(t) \rangle=0$ and $\langle W_{i}^{r}(t)W_{j}^{r'}(t') \rangle=\delta_{ij} \delta_{rr'} {\rm min}\{t,t'\}$, where the $r$ and $r'$ indices denote Descartes components.
It should be noted that the vector equation (\ref{LLG}) was replaced by two scalar equations (\ref{LLGloc1})-(\ref{LLGloc2}), since the rotation of the spin vector is always perpendicular to the direction of the spin.

During the numerical solution of equations (\ref{LLGloc1})-(\ref{LLGloc2}), the spins are rotated in sufficiently small time steps, and the components of the effective field $\frac{\partial E}{\partial \beta_{1i}},\frac{\partial E}{\partial \beta_{2i}}$ are recalculated in the new spin configuration. For the calculation of these derivatives, the band energy $E_{band}$ from density functional theory was used, defined as the single-particle grand canonical potential at zero temperature,
\begin{eqnarray}
E_{band}=\sum_{i} \varepsilon_{i} - \varepsilon_{F} N = -\int_{-\infty}^{\varepsilon_{F}}N(\varepsilon) \rmd\varepsilon
\: ,
\end{eqnarray}
where the sum goes over the occupied Kohn-Sham states and $N(\varepsilon) = \int_{-\infty}^{\varepsilon} n(\varepsilon') d\varepsilon'$ is the integrated density of states. According to the magnetic force theorem\cite{Liechtenstein,Jansen}, $E_{band}$ is a suitable alternative for the total energy if the energy differences are only calculated in lowest order of the rotation angles. The Lloyd formula\cite{Lloyd} connects the integrated density of states and the matrix of the scattering path operator (SPO) $\boldsymbol{\tau}(\varepsilon)$ within the Korringa-Kohn-Rostoker method as
\begin{eqnarray}
N(\varepsilon)&=&N_{0}(\varepsilon)+\Delta N(\varepsilon),
\\
\Delta N(\varepsilon)&=&\frac{1}{\pi}{\rm Im} \ln \det \boldsymbol{\tau}(\varepsilon) \:,
\end{eqnarray}
where $N_{0}(\varepsilon)$ is the integrated density of states of a reference system, which is independent of  the spin variables. For the band energy this leads to the expression
\begin{eqnarray}
\Delta E_{band} = -\frac{1}{\pi}\int_{-\infty}^{\varepsilon_{F}}{\rm Im} \ln \det \boldsymbol{\tau}(\varepsilon) \rmd\varepsilon = -\frac{1}{\pi} \int_{-\infty}^{\varepsilon_{F}}{\rm Im} \Tr \ln \boldsymbol{\tau}(\varepsilon) \rmd\varepsilon.
\end{eqnarray}

The Kohn-Sham effective potential  $V_{KS}$ and the exchange field $\boldsymbol{B}_{xc}$ of the system are determined by solving the Kohn-Sham-Dirac equation\cite{Kohn-Sham,Eschrig} of density functional theory in the local spin density approximation (LSDA) and using the atomic sphere approximation (ASA). In order to find the magnetic ground state the method described in \cite{Balogh2012} has been applied.

Within the LSDA and the ASA,  the exchange-correlation field $\boldsymbol{B}_{i,xc}$ at site $i$ and the corresponding spin magnetic moment $\boldsymbol{M}_{i}$,
\begin{eqnarray}
\boldsymbol{M}_{i}= -\frac{1}{\pi}\int_{-\infty}^{\varepsilon_{F}}\int_{{\rm cell}\, i}{\rm Im} \Tr \left[\beta\boldsymbol{\Sigma}G(\varepsilon,\boldsymbol{r},\boldsymbol{r})\right] \rmd^{3}\boldsymbol{r}\rmd\varepsilon,\label{Mi}
\end{eqnarray}
are parallel in the ground state. In (\ref{Mi}) $G(\varepsilon,\boldsymbol{r},\boldsymbol{r})$ denotes the Green's function, $\beta$ and $\boldsymbol{\Sigma}$ are the usual 4$\times$4 Dirac matrices, while $\varepsilon_{F}$ is the Fermi energy\cite{Eschrig}.
During the spin dynamics simulations, the effective potentials and fields were kept fixed at their ground state values, while the direction of $\boldsymbol{B}_{i,xc}$ was identified with $\boldsymbol{\sigma}_{i}$, instead of using the actual magnetic moments $\boldsymbol{M}_{i}$ in their place. Although they do not remain parallel out of the ground state, we supposed that the angle between $\boldsymbol{B}_{i,xc}$  and $\boldsymbol{M}_{i}$ remains small throughout the simulations. Also it is known that the Landau-Lifshitz-Gilbert equations conserve the length of the spin vectors $|\boldsymbol{\sigma}_{i}|=1$, while the  magnitude of the spin moments  ${M}_{i}$ may change during the simulations. These longitudinal fluctuations were also neglected in our calculations, since they were expected to be small in the case of stable magnetic moments. The validity of these assumptions will be verified in section \ref{Co10Au001}.

Up to second order in the angle variables, the single-site scattering matrix at site $i$, $t^{i}$,
changes by\cite{Udvardi}
\begin{eqnarray}
\Delta (t^{i})^{-1} = \frac{\rmi}{\hbar}
 \big[\beta_{q i}\boldsymbol{e}_{q i}\boldsymbol{J},(t^{i})^{-1}\big] - \frac{1}{2\hbar^{2}}\Big[\beta_{q i}\boldsymbol{e}_{q i}
 \boldsymbol{J},\big[\beta_{q' i}\boldsymbol{e}_{q' i}\boldsymbol{J},(t^{i})^{-1}\big]\Big],
\end{eqnarray}
when  $\boldsymbol{B}_{i,xc}$ is rotated around axis $\boldsymbol{e}_{q i}$ by angle $\beta_{q i}$ ($q=1, 2$, see figure \ref{angles}). Here $\boldsymbol{J}$ denotes the matrix of the total angular momentum operator, $[A,B]$ denotes the commutator of matrices $A$ and $B$, and a sum over the same indices ($q,q'$) has to be performed.
Using the Lloyd formula, the first and second derivatives of the band energy with respect to the angle variables can be expressed as\cite{Udvardi}
\begin{eqnarray}
\frac{\partial E_{band}}{\partial \beta_{q i}}=\frac{1}{\pi} \int_{-\infty}^{\varepsilon_{F}} {\rm Im} \Tr  \Big\{\frac{\rmi}{\hbar}\big[\boldsymbol{e}_{q i}\boldsymbol{J},(t^{i})^{-1}\big]\tau^{ii}\Big\}\rmd\varepsilon,\label{efffield}
\end{eqnarray}
\begin{eqnarray}
\frac{\partial^{2} E_{band}}{\partial \beta_{q i}\partial \beta_{q' j}}&=& \frac{1}{\pi} \int_{-\infty}^{\varepsilon_{F}} {\rm Im} \frac{1}{\hbar^{2}} \Tr \Bigg(\big[\boldsymbol{e}_{q i}\boldsymbol{J},(t^{i})^{-1}\big]\tau^{ij} \big[\boldsymbol{e}_{q' j}\boldsymbol{J},(t^{j})^{-1}\big]\tau^{ji} \nonumber
\\
&&- \delta_{ij} \frac{1}{2}\Big\{\Big[\boldsymbol{e}_{q i}\boldsymbol{J},\big[\boldsymbol{e}_{q' i}\boldsymbol{J},(t^{i})^{-1}\big]\Big]\nonumber
\\
&&+\Big[\boldsymbol{e}_{q' i}\boldsymbol{J},\big[\boldsymbol{e}_{q i}\boldsymbol{J},(t^{i})^{-1}\big]\Big]\Big\}\tau^{ii} \Bigg)\rmd\varepsilon\label{secder} \: .
\end{eqnarray}

The first derivative appears explicitly in the Landau-Lifshitz-Gilbert equations (\ref{LLGloc1})-(\ref{LLGloc2}), when using a local coordinate system. The second derivatives will be used in the one-step numerical integration scheme detailed in the next section. It is worth mentioning that the second derivatives for a ferromagnetic configuration are related to the exchange coupling tensor and (\ref{secder}) simplifies to the Liechtenstein formula\cite{Liechtenstein} in the nonrelativistic case.

\section{Numerical integration algorithms\label{numint}}

Equations (\ref{LLGloc1})-(\ref{LLGloc2}) describe the motion of the spins in the local coordinate system. As in each time step the calculation of the effective field is quite demanding, a numerical integration scheme is needed to solve the system of stochastic differential equations which can be used with a relatively large time step. Three numerical integration schemes were employed for calculating the next spin configuration $\boldsymbol{\sigma}_{i}(t_{n+1})$ at time $t_{n+1}=t_{n}+\Delta t$ from the current spin configuration $\boldsymbol{\sigma}_{i}(t_{n})$, using small rotations $\Delta \beta_{1i}, \Delta \beta_{2i}$ and the derivatives (\ref{efffield}) and (\ref{secder}). The computational details of these integration schemes are given in \ref{numintschemes}, here only the basic features of the algorithms are summarized.

Conserving the length of the spin vectors is an important symmetry of the equations, since during the calculation of the effective field the spin vectors are supposed to be normalized. Unfortunately, the Heun method, which is the most widely used numerical scheme to solve the stochastic Landau-Lifshitz-Gilbert equation\cite{Palacios,Skubic}, does not fulfill this requirement. Recently Mentink \etal\cite{Mentink} have proposed a method which does conserve the magnitude of the spins.
Modified for the local coordinate system, this algorithm can be sketched as
\begin{eqnarray}
\boldsymbol{\sigma}_{i}(t_{n}) \rightarrow \frac{\partial E}{\partial \beta_{1i}}, \frac{\partial E}{\partial \beta_{2i}} \rightarrow \boldsymbol{\tilde{\sigma}}_{i}(t_{n}) \rightarrow \frac{\partial E}{\partial \tilde{\beta}_{1i}},\frac{\partial E}{\partial \tilde{\beta}_{2i}} \rightarrow \boldsymbol{\sigma}_{i}(t_{n+1}),
\end{eqnarray}
where $\boldsymbol{\tilde{\sigma}}_{i}(t_{n})$ is a first approximation for $\boldsymbol{\sigma}_{i}\big(\frac{1}{2}(t_{n}+t_{n+1})\big)$. This is a two-step numerical integration scheme, since the derivatives have to be calculated for two different spin configurations, $\boldsymbol{\sigma}_{i}(t_{n})$ and $\boldsymbol{\tilde{\sigma}}_{i}(t_{n})$. Since the most time-consuming part of the ab initio simulation is the calculation of the scattering path operator $\boldsymbol{\tau}$, a method  would be more preferable where the effective fields are calculated only once for a time step, but the scheme has similar stability and convergence properties to the above solver.

Therefore we propose the one-step scheme with the algorithm
\begin{eqnarray}
\boldsymbol{\sigma}_{i}(t_{n}) \rightarrow \frac{\partial E}{\partial \beta_{qi}}, \frac{\partial^{2} E_{band}}{\partial \beta_{q i}\partial \beta_{q' j}} \rightarrow \boldsymbol{\sigma}_{i}(t_{n+1}),
\end{eqnarray}
where it is necessary to evaluate the second derivatives of the energy.  
Here the determination of the new configuration from the derivatives is more complex than in the two-step scheme, see \ref{numintschemes}. Nevertheless,  the computational time of a time step for the one-step scheme is still much smaller than for the two-step scheme.

We also examined the simplified one-step scheme with the algorithm
\begin{eqnarray}
\boldsymbol{\sigma}_{i}(t_{n}) \rightarrow \frac{\partial E}{\partial \beta_{1i}}, \frac{\partial E}{\partial \beta_{2i}} \rightarrow \boldsymbol{\sigma}_{i}(t_{n+1}),
\end{eqnarray}
which is based on the Euler method. This method exhibits the beneficial properties of both the one-step and two-step schemes: the effective fields have to be calculated only once for each time step and the calculation of the new spin configuration from the effective field has a simpler form than in the one-step scheme.

As given in \ref{numintschemes}, all three methods have weak order of convergence $\delta=1$, but they have different stability properties. In section \ref{model} it will be demonstrated that the simplified one-step scheme is much less stable than the other two methods, therefore a significantly smaller time step is necessary, which considerably increases the length of the simulation.

\section{Applications to a one-dimensional Heisenberg chain\label{model}}

Before implementing the numerical solver in the embedded cluster Korringa-Kohn-Rostoker method\cite{Lazarovits}, the different schemes discussed in section \ref{numint} were compared for the case of a one-dimensional classical Heisenberg chain, described by the Hamiltonian
\begin{eqnarray}
E=J\sum_{i=1}^{N-1}\boldsymbol{\sigma}_{i}\boldsymbol{\sigma}_{i+1},\label{modelHam1}
\end{eqnarray}
where $N$ is the number of spins, ferromagnetic coupling  $J<0$ was considered between the nearest neighbours and free boundary conditions were used. The expectation value of the energy as a function of temperature can be explicitly given as\cite{Shubin,Fisher}
\begin{eqnarray} \label{ETanal}
\langle E \rangle (T)=(N-1) JL\left(\frac{J}{k_{B}T}\right),
\end{eqnarray}
where $L(x) = \frac{1}{x} - \coth(x)$ is the Langevin function multiplied by $-1$.
The average of the square of the magnetization can be calculated as\cite{Fisher}
\begin{eqnarray}
\langle \boldsymbol{M}^{2} \rangle&=&\left\langle \left( \frac{1}{N}\sum_{i}\boldsymbol{M}_{i}\right)^{2} \right\rangle\nonumber
\\
&=&\frac{\mu^{2}}{N^{2}}\Bigg[N\frac{1+L(\frac{J}{k_BT})}{1-L(\frac{J}{k_BT})} - 2L\bigg(\frac{J}{k_BT}\bigg)\frac{1-L(\frac{J}{k_BT})^{N}}{\big[1-L(\frac{J}{k_BT})\big]^2}\Bigg],\label{M2Tanal}
\end{eqnarray}
where $\mu$ is the size of the atomic magnetic moment. For the model Hamiltonian (\ref{modelHam1})  $\mu=1$, while a value of $\mu \ne 1$ will be fitted to the ab initio results in section \ref{Co10Au001}.

Since in this case the energy is known as a function of the spin vectors in the global coordinate system, the global two-step scheme proposed in \cite{Mentink} can be compared to the methods applied in the local coordinate system. Explicit expressions for the first and second derivatives of the energy in the local coordinate system are given in \ref{appendix2}.

\begin{figure}
\centering
\includegraphics[width=\columnwidth]{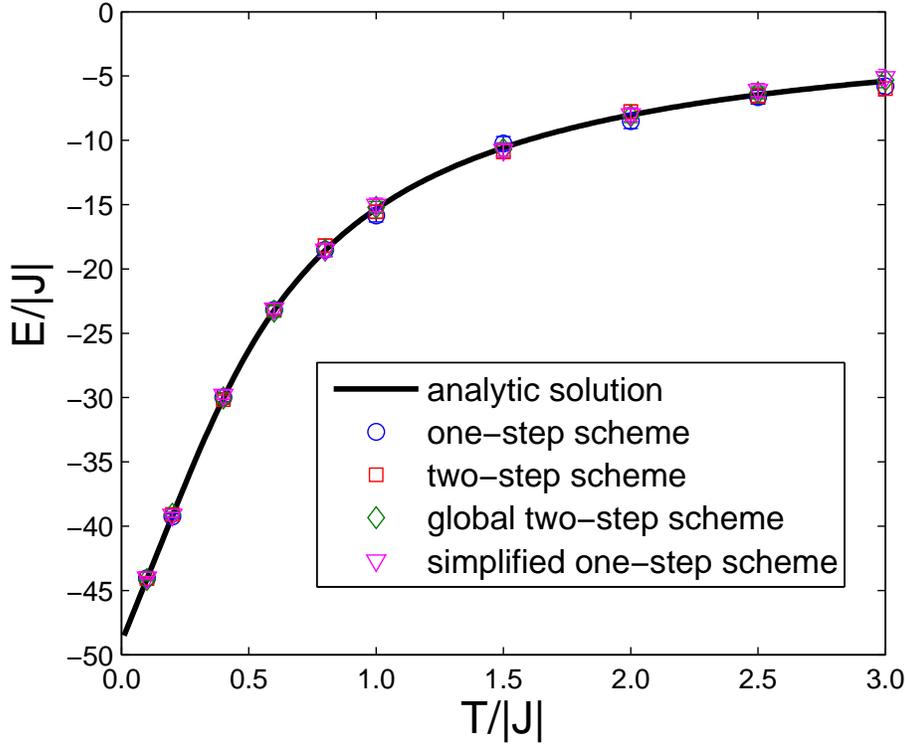}
\caption{Statistical average of the energy of a linear chain of $N=50$ spins as a function of the temperature obtained using  different numerical schemes. The units of $J=-1$ and $k_{B}=1$ are used, with the damping value $\alpha=0.05$. The expectation value is calculated by running the simulation for 500000 time units, and averaging the value of the energy at the last time step over 200 different realizations, that is different seeds of the random number generator. The (very small) error bars denote the 95\% confidence intervals, see \ref{confidence}. The time step was $\rmd t=0.05$ for the first three schemes, and $\rmd t=0.001$ in the case of the simplified one-step scheme.}
\label{ET}
\end{figure}

\begin{figure}
\centering
\includegraphics[width=\columnwidth]{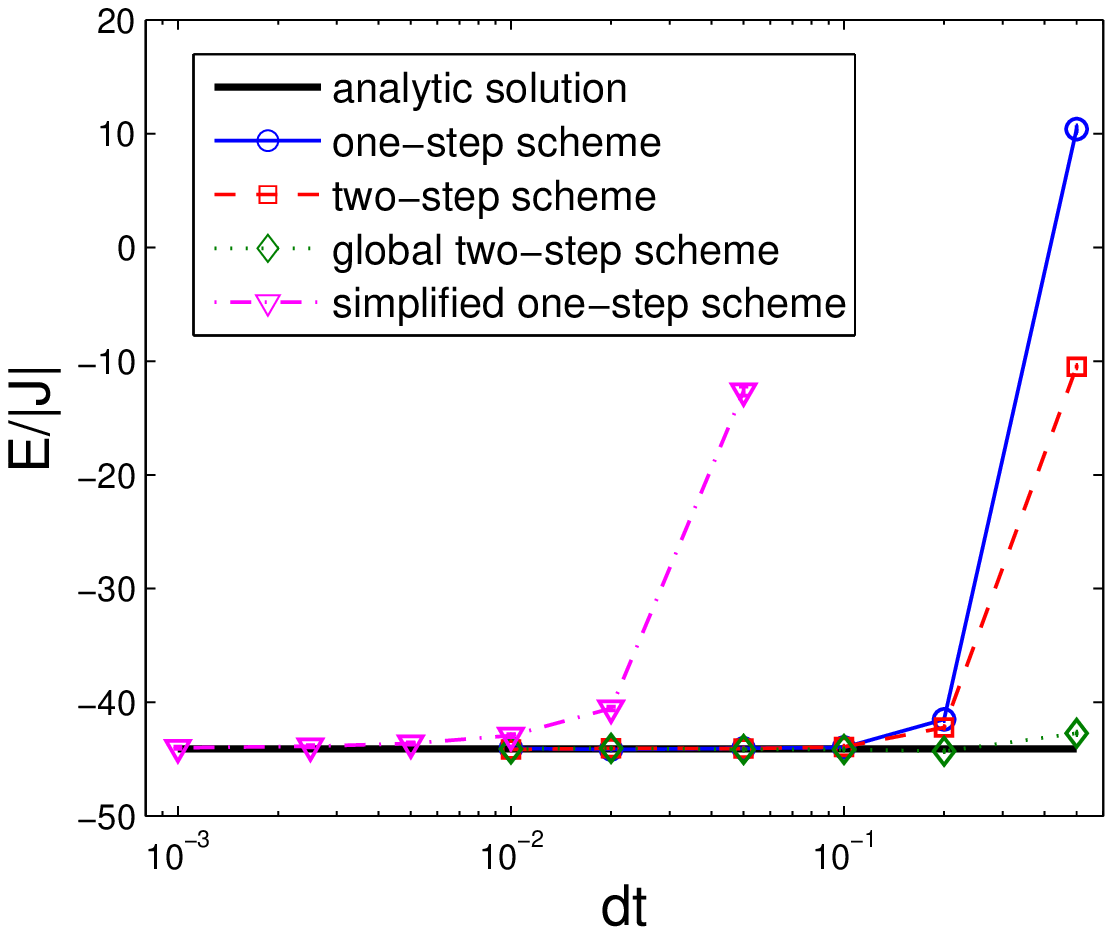}
\caption{Statistical average of the energy of a linear chain of $N=50$ spins  as a function of the time step obtained using the different numerical schemes. The units of $J=-1$ and $k_{B}=1$ are used and the temperature was fixed to $T=0.1$, with the damping value $\alpha=0.05$. The expectation value is calculated by running the simulation for 500000 time units, and averaging the value of the energy at the last time step over 200 different realizations (different seeds of the random number generator). The small error bars denote the 95\% confidence intervals.}
\label{Edt}
\end{figure}

For the simulations a ferromagnetic system with $J=-1$ was chosen, and the mean energy was calculated as a function of temperature for each of the numerical schemes. As can be seen in figure \ref{ET}, all the proposed methods give results which are in relatively good agreement with the analytic solution. In order to reach appropriately low error values, the simplified one-step scheme requires a much smaller time step than the other methods. This can also be seen in figure \ref{Edt}, where the mean energy is depicted at a given temperature, as a function of the size of the time step. The one-step and two-step methods have similar stability properties, both of them being in agreement with the analytic result for the expectation value of the energy up to time steps $\rmd t \approx 0.1$. On the contrary, the simplified one-step scheme requires an about $100$ times smaller time step. The most efficient method in this case is the two-step scheme compiled in the global coordinate system, where one can use about $5$ times larger time steps as in the one-step and two-step schemes using the local coordinate system. However, this approach does not fit the requirements of the embedded cluster Korringa-Kohn-Rostoker method.

We thus conclude that the most effective numerical method for the ab initio calculations is the one-step scheme, as it has the same stability properties as the two-step scheme, but requires less computational capacity since at each time step the derivatives have to be calculated only for a single spin configuration as discussed in section \ref{numint}.

In order to implement the one-step method in ab initio calculations an appropriate time scale for the magnetic processes must be determined. In the case of a simple Heisenberg model, the only parameter is the exchange coupling $J$ with the corresponding time scale $1/|J|$. As it is demonstrated in figure \ref{Edt} the one-step scheme remains stable up to time steps as large as $5-10\%$ of this time scale. In the ab initio calculations, the magnitudes of the atomic magnetic moments, the interactions between the spins and the effect of the underlying lattice all influence the time scales of the system, therefore it is important to determine them before starting the simulations.

For the simulations in section \ref{Co10Au001}, the appropriate time scales were determined from the $\omega_{k}$ frequencies of the normal modes of the spin system without damping, close to the ground state. The method for determining these frequencies is given in \ref{normalmodes}. The largest frequency corresponds to the smallest characteristic time period, which in turn determines the correct time step in the simulation. On the other hand, the smallest frequency related to the largest time scale helps in determining the length of the simulation. For example, the angular frequencies for a simple Heisenberg chain with periodic boundary conditions will be distributed between $0$ and $4|J|$, with $\omega_{k}=-2J\big(1-\cos{2\pi\frac{k}{N}}\big)$ for $k=0, \dots, N-1$. Comparing this to figure \ref{Edt}, we can conclude that the one-step scheme remains stable up to time steps $\Delta t \approx 0.4 \, \omega_{max}^{-1}$. The relaxation processes due to the damping $\alpha$ also influence the time scales, but in the case of $\alpha \ll 1$ which is usually a good assumption for stable magnetic moments, the relaxation processes are significantly slower than the oscillations.

\section{Application to Co/Au($001$)\label{Co10Au001}}

\begin{figure}
\centering
\includegraphics[width=\columnwidth]{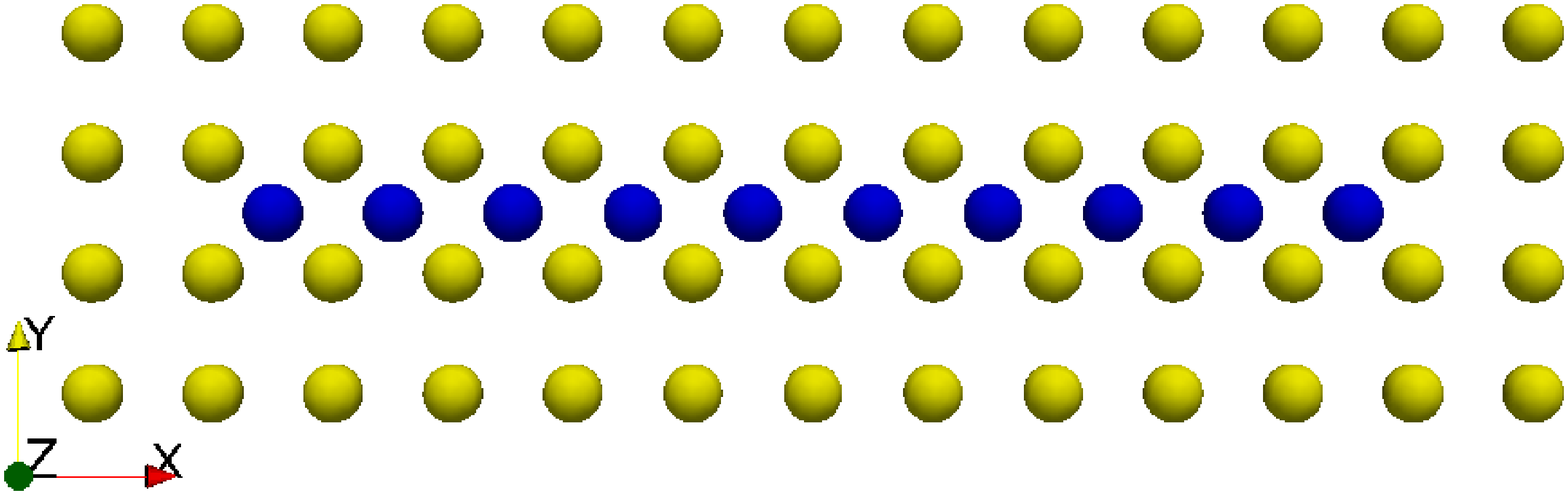}
\includegraphics[width=0.8\columnwidth]{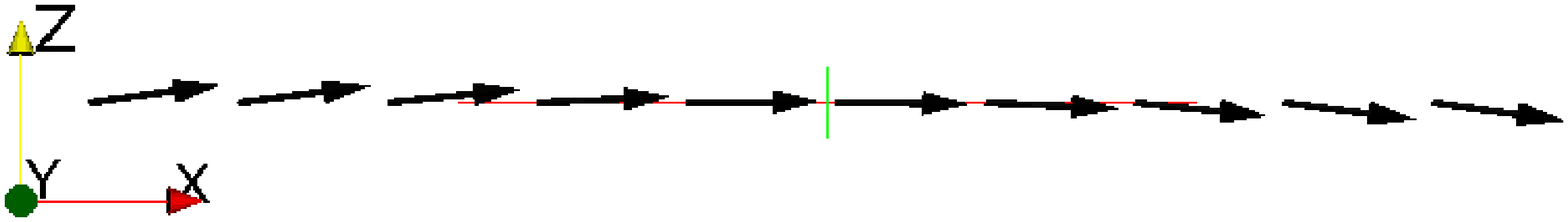}
\caption{Top view of ten cobalt atoms (blue circles) forming a linear chain above Au($001$) surface (gold circles). The ground state configuration of the spin vectors of the cobalt atoms is also sketched.}
\label{cobalt}
\end{figure}

For the ab initio simulations we chose a linear chain of Co atoms deposited in the hollow positions above Au(001) surface, see figure \ref{cobalt}. Lattice relaxations were not included in the calculations, that is both the Au surface layers and the
deposited Co atoms preserved the positions of the Au bulk fcc lattice.
The magnetic ground state configuration and the corresponding effective potentials and exchange fields have been determined self-consistently by using the method described in \cite{Balogh2012}. The obtained ground state spin configuration is also depicted in figure \ref{cobalt}. As mentioned before, the effective potential obtained for the ground state is kept constant during the simulations, while only the direction of the exchange field is changed according to the Landau-Lifshitz-Gilbert equations.

If the system can be described by a model Hamiltonian (\ref{modelHam1}), then the $J$ exchange interaction can be calculated from the second derivatives (\ref{secder}) in the ground state, by using equations (\ref{Jcalc1})-(\ref{Jcalc2}) in \ref{appendix2}. Due to relativistic effects, in particular
spin-orbit coupling, the second derivatives (\ref{Jcalc1})-(\ref{Jcalc2}) give different $J$ values even for the same pair of atoms, therefore we averaged them to obtain a reasonable estimate for the scalar coupling. The calculated nearest-neighbour exchange parameters took values between $-3.16$ and $-4.47$ mRyd, being enhanced at the ends of the cluster, with an average value of $J_{av}=-3.58$ mRyd. These values are remarkably smaller than the ones reported by Tung \etal\cite{Tung2011B} ($-11.5$ mRyd) and by T\"ows \etal\cite{Tows}($\approx-13$ mRyd at $T=0$) for free-standing chains.
The main reason for this difference is that the intersite distance in the free-standing chains is smaller than that determined by the lattice constant of the fcc lattice of Au we used in our calculations.

The interactions between the next-nearest neighbours appeared to be ferromagnetic, but about ten times smaller than for the nearest neighbours, while between the third-nearest neighbours an antiferromagnetic coupling was found, all of these in good agreement with earlier results\cite{Tung2011B,Tows}. Contrary to the ferromagnetic state reported in these works\cite{Tung2011B,Tows}, we obtained a ground state resembling a spin spiral, which we attribute to the appearance of Dzyaloshinsky-Moriya interactions.
Since the system has a mirror symmetry with respect to the $x\!-\!z$ plane as shown in figure \ref{cobalt}, it can be shown\cite{Moriya} that the Dzyaloshinsky-Moriya vectors are
parallel to the $y$ axis, leading to a spin spiral in the $x\!-\!z$ plane.
Note that the Dzyaloshinsky-Moriya interactions only arise due to breaking of inversion symmetry
in the presence of the substrate, therefore they do not appear for infinite free-standing chains\cite{Tung2011B,Tows}. It can also be inferred from figure \ref{cobalt} that the chain direction ($x$) is an easy magnetization axis, just as it was
found for Cu($001$) surface\cite{Hong,Tung2011}.

\begin{figure}
\centering
\includegraphics[width=\columnwidth]{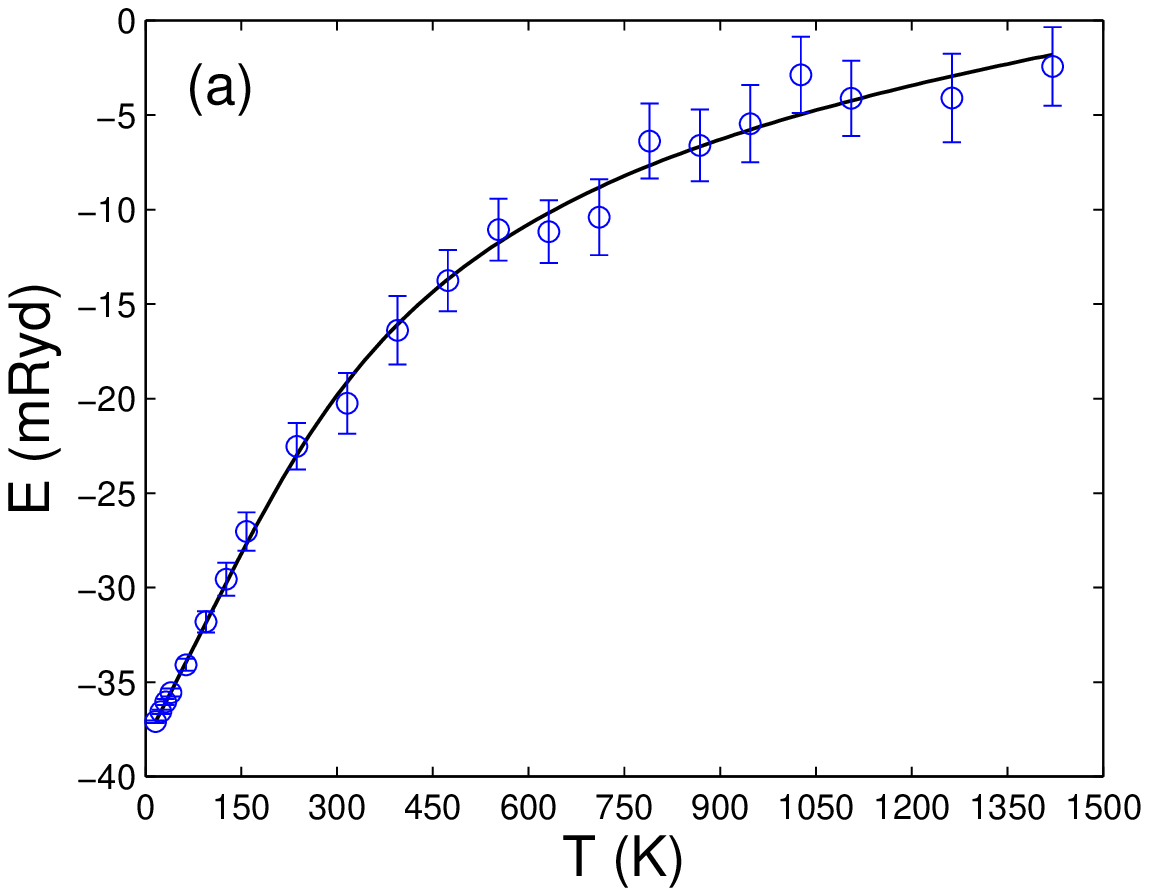}
\includegraphics[width=\columnwidth]{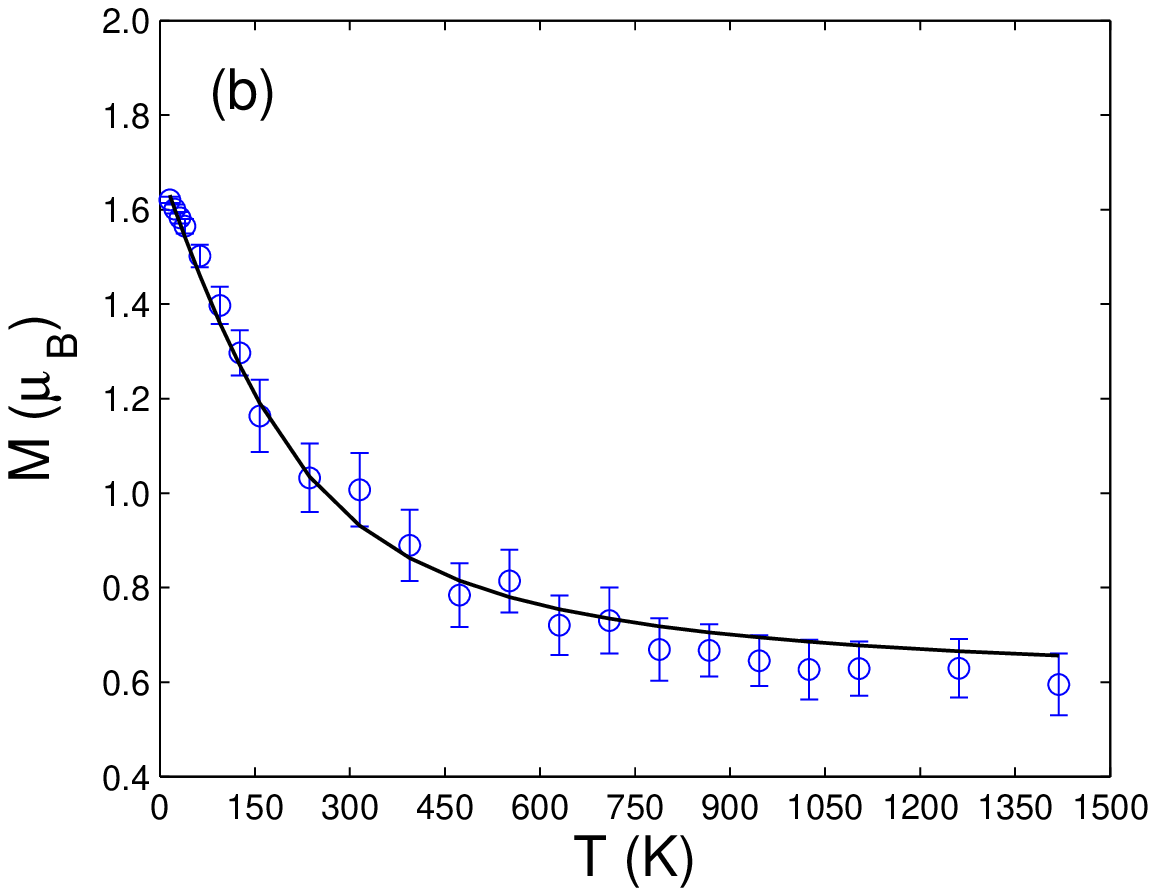}
\caption{The mean value of the energy (a) and of the magnetic moments (b) of a chain of ten Co atoms on Au(001) as a function of temperature. The circles correspond to the simulation results, the solid lines are the fitted curves using equations (\ref{ETanal}) and (\ref{M2Tanal}), respectively. The quantity $M$ in panel (b) is calculated as $M=\sqrt{\langle \boldsymbol{M}^2 \rangle}$. The expectation values are calculated by running the simulations for 100000 time units, and taking the average  at the last time step over 50 different realizations, that is different seeds of the random number generator. The error bars denote the 95\% confidence intervals. The time unit is $1\frac{\hbar}{{\rm Ryd}}=48.5$~as, with the time step being $5$ time units. The time step was determined by calculating the normal modes of the system as discussed in section \ref{model}, yielding a maximal frequency  $\omega_{max}=8.27\,\frac{{\rm mRyd}}{\hbar}$. The value of the damping parameter was $\alpha=0.05$.}
\label{Co10EMT}
\end{figure}

Firstly the thermal behaviour of the spin system was compared to the model Hamiltonian (\ref{modelHam1}).
In figure~\ref{Co10EMT} the mean value of the energy and the magnetic moment of the system, defined as $M=\sqrt{\langle \boldsymbol{M}^2 \rangle}$ with $ \boldsymbol{M}=\frac{1}{N} \sum_{i=1}^{N} \boldsymbol{M}_{i} $, are shown as a function of temperature.
The mean value for the energy was fitted using the analytic expression (\ref{ETanal}),
yielding the value $J=-3.64\pm0.24$ mRyd, which is close to the average value of the scalar coupling
coefficients between the spins, $J_{av}=-3.58$ mRyd, calculated directly before.
Using the previously fitted exchange coupling $J$, the mean magnetic moment from the simulation results in figure \ref{Co10EMT} was fitted using (\ref{M2Tanal}), resulting in the value $\mu=1.694\pm0.006$ $\mu_{B}$.
The ab initio calculations (equation (\ref{Mi})) yielded magnetic moments between $1.656$ $\mu_{B}$ and $1.689$ $\mu_{B}$, with the average value of $\mu_{av}=1.670$ $\mu_{B}$, in agreement with the above fitted value.

\begin{figure}
\centering
\includegraphics[width=\columnwidth]{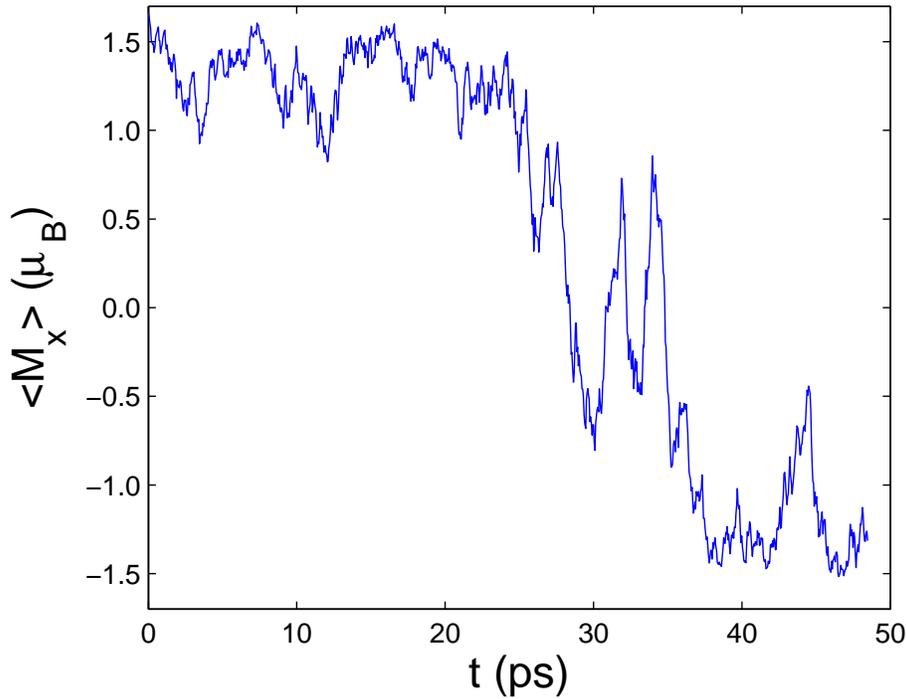}
\caption{The mean value of the $x$ component of the average magnetic moment as a function of simulation time, when the simulation is started from a configuration when all spins point towards the positive $x$ axis. The temperature was $T=78.8$~K, the damping $\alpha=0.05$.}
\label{flip}
\end{figure}

Ignoring chirality effects due to the Dzyaloshinsky-Moriya interactions, the magnetic anisotropy prefers all spins pointing parallel to the $x$ direction. Since the system is invariant under time reversal, it has two degenerate ground states, namely all spins pointing towards either the positive or the negative $x$ direction. Due to the energy barrier between these two states, the system freezes in one of these ground states at $T=0$. However, at finite temperature, the system will be continuously switching between these degenerate states.  Such a switching process is presented  in figure~\ref{flip} showing the temporal variation of the $x$ component of the average spin of the Co chain at  $T=78.8$ K.

During the switching process the spin system gets relatively far from the ground state configuration, therefore it was tempting to verify the assumption made at the end of section \ref{efff}, namely that the deviation between
the directions of the exchange fields and the magnetic moments remains small. In each time step the
direction of the exchange field $\{\boldsymbol{\sigma}_i\}$ was compared to the orientation of the calculated spin magnetic moment and it was found that the angle between $\boldsymbol{B}_{i,xc}$ and $\boldsymbol{M}_{i}$ was never larger than $3^{\circ}$.
Moreover, the magnitudes of $\boldsymbol{M}_{i}$ fluctuated within just a $\pm 2\%$ wide range around the corresponding ground state value, occasionally reaching values up to $\pm 5\%$.
Consequently, we concluded that, at least in case of stable magnetic moments, the magnetic force theorem can be applied in ab initio spin dynamics simulations.

Calculating the switching time between the two ground states gives information about the anisotropy energy of the system. It is expected that the switching time, $\tau_{sw}$, follows the Arrhenius-N\'eel law\cite{Neel} as the function of temperature,
\begin{eqnarray}
\tau_{sw}=\tau_{0}e^{\frac{\Delta E}{k_{B}T}},\label{Neel}
\end{eqnarray}
where $\tau_0$ and $\Delta E$ are appropriate constants. 

\begin{figure}
\centering
\includegraphics[width=\columnwidth]{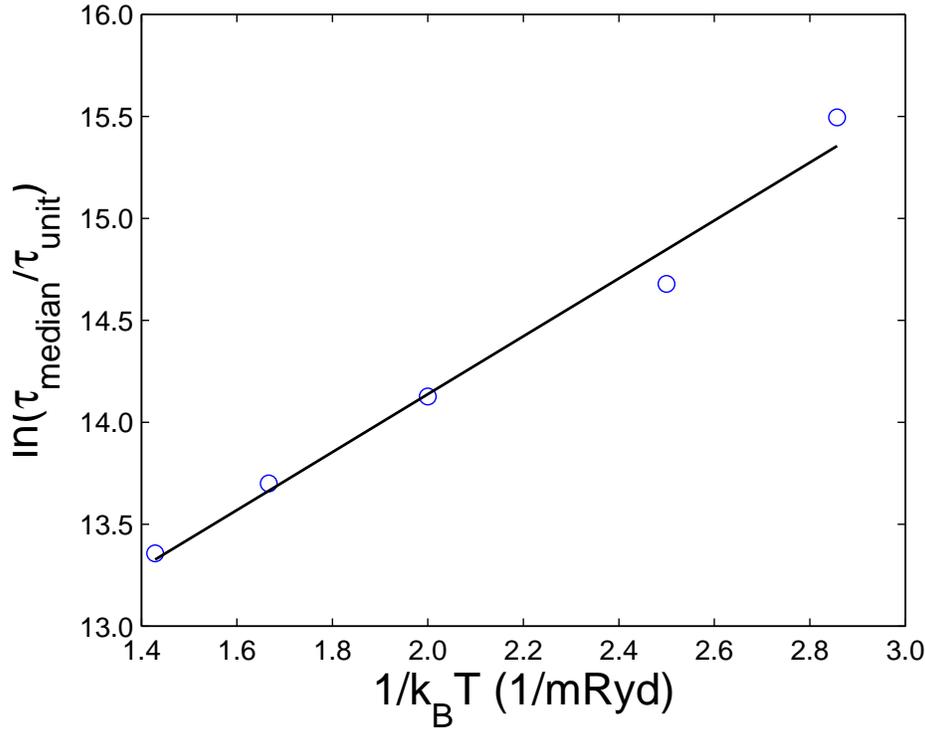}
\caption{The median value of the switching time as a function of the inverse temperature (open circles), along with the fitted linear curve (solid line). The time unit is $\tau_{unit}=48.5$~as. The median value $\tau_{median}$ was obtained from 50 independent runs at a given temperature. The value of the damping parameter was $\alpha=0.05$.}
\label{switch}
\end{figure}

The switching time from the simulations was determined by starting the simulation from the $+x$ direction and taking the first time when $\left\langle M_{x} \right\rangle <-1.0$, which is relatively close to the state when the spins point towards the $-x$ direction as can be inferred from figure \ref{flip}. Performing simulations for several different realizations of the noise,
the median value of the switching times, $\tau_{median}$, was taken at a given temperature, since calculating $\tau_{median}$ instead of the average of the switching times requires less computation time: one has to take the middle value of the flipping times, so the maximal simulation time corresponds to the time interval for which half of the realizations displays a flipping. It was assumed that the switching time has an exponential distribution with expectation value $\tau_{sw}$, and in this case the simple proportionality $\tau_{median}=\ln{2}\,\tau_{sw}$ holds, therefore $\tau_{median}$ also follows (\ref{Neel}), only with a different $\tau_{0}$. It can be seen in figure \ref{switch} that this is indeed the case: $\ln{\tau_{median}}$ is approximately a linear function of $\frac{1}{k_{B}T}$.

Related to the switching process, we compared our results from ab initio spin dynamics simulations to that from the simple model Hamiltonian
\begin{eqnarray}
E=J\sum_{i=1}^{N-1}\boldsymbol{\sigma}_{i}\boldsymbol{\sigma}_{i+1}+K\sum_{i=1}^{N}\sigma_{ix}^{2},\label{modelHam}
\end{eqnarray}
with $N=10$, $J=-3.6$ mRyd, and in the dynamical equations (\ref{LLGloc1}) and (\ref{LLGloc2}) we used $M_{i}=1.67$ $\mu_{B}$ at every site. 
The uniaxial anisotropy supposed in the above model is just an approximation, since the symmetry of the system implies in fact biaxial anisotropy. Indeed, ab initio calculations in terms of the magnetic force theorem resulted in different energies for magnetizations along the $x$, $y$ and $z$ directions: $\frac{1}{N}\big(E_{x}-E_{y}\big)=-0.26$ mRyd and $\frac{1}{N}\big(E_{x}-E_{z}\big)=-0.17$ mRyd. It turned out that the value $K=-0.24$ mRyd was the most appropriate for the model calculations. A comparable value $K=-0.09$ mRyd was found for an infinite Co chain on Cu($001$)\cite{Tung2011}.

\begin{table}
\caption{The parameters of a linear function fitted to the $\ln(\tau_{median}/\tau_{unit})$ data versus $\frac{1}{k_{B}T}$ as obtained from the ab initio simulations, see figure \ref{switch}, and from the spin model, equation (\ref{modelHam}),
with $J=-3.6$~mRyd and $K=-0.24$~mRyd. \label{fit}}
\begin{indented}
\item[]\begin{tabular}{@{}ccc}
\br
           &  \text{ab initio} &      \text{model} \\
\mr
 ln$(\tau_{0}/\tau_{unit})$ & 11.30$\pm$0.24
 & 11.26$\pm$0.18
 \\
\mr
     $\Delta E$ (mRyd) & 1.42$\pm$0.11
 &  1.46$\pm$0.08 \\
\br
\end{tabular}
\end{indented}
\end{table}

\begin{figure}
\centering
\includegraphics[width=\columnwidth]{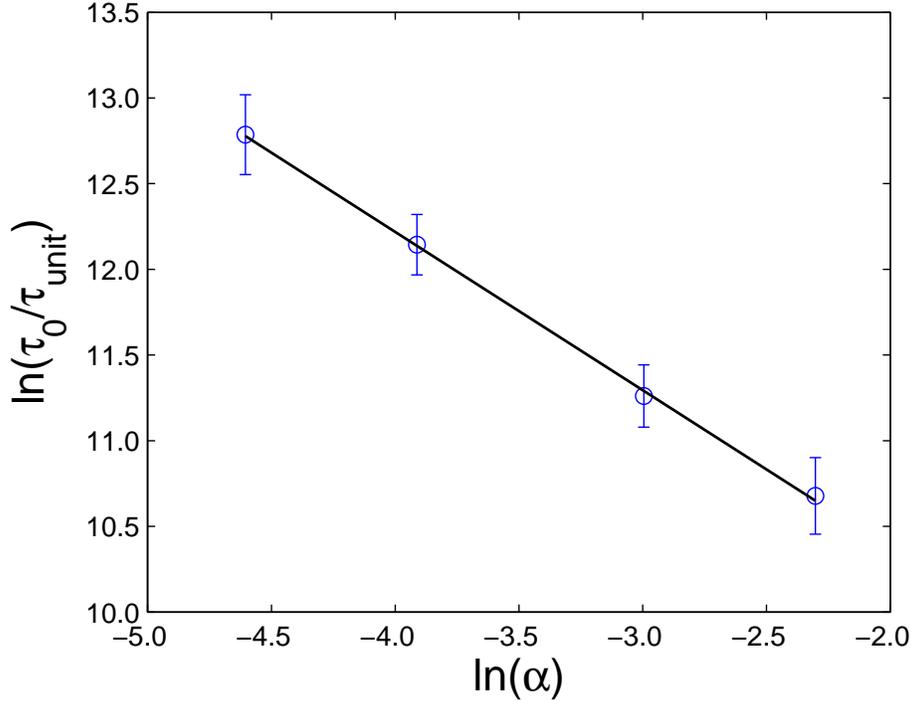}
\caption{The dependence of the parameter $\tau_{0}$ in (\ref{Neel}) on the Gilbert damping constant $\alpha$. Open circles represent the intercept values of the curves fitted to the simulation results as in figure \ref{switch}, but for different values of $\alpha$, while the error bars show the error of these fitting parameters. The solid line displays a best fit linear function to $\ln(\tau_{0}/\tau_{unit})$ as a function of $\ln(\alpha)$, indicating a power law dependence of $\tau_{0}$ on $\alpha$.}
\label{alpha}
\end{figure}

By using the spin model (\ref{modelHam}) the switching times were calculated in the same way as in the ab initio simulations. The above model parameters ensured a linear dependence of $\ln{\tau_{median}}$ on the inverse temperature with parameters coinciding almost precisely with those from the ab initio calculations, see Table \ref{fit}.
Therefore we conclude that the investigated system can be well described by the model Hamiltonian (\ref{modelHam}). Flipping times for the same model Hamiltonian were examined in detail in \cite{Bauer}, where an asymptotic expression is given for $\Delta E$ for the cases $N \ll L_{DW}$ and $N \gg L_{DW}$, with $L_{DW}=2\sqrt{J/K}$ being the domain wall width in the chain. With $N=10$ and $L_{DW}=7.75$, our model calculation falls in the intermediate regime.

Finally we examined the dependence of the fitting parameters on the damping parameter $\alpha$. The simulations using the model Hamiltonian  (\ref{modelHam}) were carried out for the values $\alpha=0.01,0.02,0.05$ and $0.1$, and a linear dependence was supposed between $\ln(\tau_{median}/\tau_{unit})$ and $\frac{1}{k_{B}T}$. It was found that the slope of the curve $\Delta E$ does not depend on $\alpha$ as can be expected since this quantity is determined by the free energy landscape and it is fairly independent of the dynamical behaviour. On the other hand, the intercept value $\tau_{0}$ does depend on the damping, with the power law dependence $\tau_{0} \propto \alpha^{x}$ as indicated in figure \ref{alpha}. The exponent of the power law was found to be $x=-0.92\pm0.12$, showing an approximate inverse proportionality between the two quantities.

\section{Summary and conclusions}

We proposed a new method to study  the magnetism of small clusters at finite temperature. The method is based on the quasiclassical stochastic Landau-Lifshitz-Gilbert dynamics, where the effective field $\boldsymbol{B}_{i}^{eff}$ acting on the spin vectors is directly determined from ab initio calculations during the numerical solution of the dynamical equations instead of using an effective spin Hamiltonian. For this purpose we employed the torque method as implemented within the embedded cluster Korringa-Kohn-Rostoker multiple scattering method. During the time evolution the classical spin vectors $\boldsymbol{\sigma}_{i}$ were identified with the direction of the exchange-correlation magnetic field $\boldsymbol{B}_{i,xc}$ at a given lattice point, and we assumed that this direction remains close to the direction of the spin magnetic moment $\boldsymbol{M}_{i}$ calculated from first principles. Furthermore, it was assumed that the magnitude of the stable moments does not vary considerably during the time evolution. In case of stable magnetic moments under investigation, these assumptions were well justified, since the angle between $\boldsymbol{B}_{i,xc}$ and $\boldsymbol{M}_{i}$ remained below $3^{\circ}$, while the relative longitudinal fluctuations did not exceed $5\%$.

Using the above first principles scheme, the stochastic Landau-Lifshitz-Gilbert equations have to be solved in the local coordinate system (the local $z$ axis is fixed  along $\boldsymbol{\sigma}_{i}$),  therefore an appropriate numerical solver had to be developed. Based on the semi-implicit method developed by Mentink \etal\cite{Mentink}, we proposed three numerical schemes, which were tested for a one-dimensional Heisenberg chain with nearest-neighbour interactions. It was found that although all three methods are able the reproduce the analytic results for the mean energy of the system as a function of temperature, the one-step scheme is the most preferable, since there a $100$ times larger time step can be used than in the simplified one-step scheme and, at each time step, the derivatives of the energy have to be calculated only for a single spin configuration, contrary to the two-step scheme, where they have to be calculated for two different spin configurations.

This method was applied to a linear chain of ten Co atoms deposited on Au($001$) surface. In agreement with recent results on infinite Co chains, either free-standing or supported by Cu(001)\cite{Hong,Tung2011,Tung2011B,Tows}, we found that this system is governed by strong ferromagnetic exchange couplings with an easy magnetization axis along the chain direction. Nonetheless, due to the presence of Dzyaloshinsky-Moriya interactions, we obtained a ground state with slightly tilted spins, resembling a spin spiral. Performing finite-temperature simulations we found that the mean energy and the mean magnetization can be approximated with a high accuracy by using a ferromagnetic Heisenberg model with suitable parameters. We demonstrated that the switching process between the degenerate ground states, with the spins pointing towards the $+x$ or the $-x$ directions, can be well described by adding an on-site anisotropy term to the model Hamiltonian.

We plan to apply the method to systems with more complex geometry where the design of an appropriate spin model is less obvious. Special interest should be devoted to the study of nanomagnets where higher order interactions may take place between the spins\cite{Heinze}. Furthermore, it is also worthwhile to extend the method by including induced magnetic moments in the calculations, although the stable spin description is not suitable for these types of atoms. Another possible extension of the method includes longitudinal spin-fluctuations by recalculating the potentials and effective fields at every temperature according to finite-temperature density functional theory\cite{Mermin,Janak}, since this may strongly influence the spin-interactions in an ab initio based spin Hamiltonian, especially at higher temperatures as shown in \cite{Tows} and \cite{Chimata}.

\ack
The authors thank Professor Ulrich Nowak for useful discussions and suggestions. Financial support was provided by the Hungarian National Research Foundation (under contracts OTKA 77771 and 84078), and in part by the European Union under FP7 Contract No. NMP3-SL-2012-281043 FEMTOSPIN. The work of LS was supported by the European Union, co-financed by the European Social Fund, in the framework of T\'AMOP 4.2.4.A/2-11-1-2012-0001 National Excellence Program.

\appendix
\setcounter{section}{1}

\section*{Appendix}

\subsection{Numerical integration schemes\label{numintschemes}}

For a thorough description of the type and order of convergence of stochastic numerical integration techniques the reader is referred to the handbook on stochastic numerical schemes\cite{Kloeden}. Here only the schemes used in this paper are described. A stochastic diffusion differential equation in one dimension has the form
\begin{eqnarray}
\rmd X(t) & = & a(X(t),t)\rmd t+b(X(t),t) \circ \rmd W(t),
\label{eq1}
\\
 X(t_{0})&=&  X_{0},
\end{eqnarray}
written in the Stratonovich form.

During the numerical procedure the exact solution $X(t)$ is approximated on the time interval $[0,T]$ by a process $Y^{\Delta t}(t)$, which is only defined at certain discrete points in time, and the largest difference between the discrete time points is $\Delta t$. Physical quantities, like the energy and magnetization of the system discussed in the paper, correspond to averages or expectation values over the trajectories. If only the expectation value of some function $g(X(t))$ of the exact solution $X(t)$ has to be approximated, the weak convergence criterion can be applied: $Y^{\Delta t}(t)$ converges to the solution $X(t)$ weakly with order $\delta>0$, if there exists a constant $C$ such that
\begin{eqnarray}
|\langle g(X(T))\rangle-\langle g(Y^{\Delta t}(T))\rangle| \leq C \Delta t^{\delta} \: ,
\end{eqnarray}
for a given set of test functions $g(x)$, where $\langle \rangle$ denotes stochastic expectation value. Numerical integration schemes can be constructed by using the stochastic Taylor expansion of the exact solution. For a theorem on calculating the weak order of convergence of a given numerical method, see p 474 of \cite{Kloeden}.

An important property of the numerical integration schemes for the stochastic Landau-Lifshitz-Gilbert equation considered in this paper is the conservation of the length of the spin vectors, which should be reflected in the numerical solver as suggested in \cite{Mentink}. If the spin vectors are known at time $t_{n}$, their value one time step later at $t_{n+1}$ can be evaluated by combining (\ref{e1e2}) and (\ref{dsigma}),
\begin{eqnarray}
\boldsymbol{\sigma}_{i}(t_{n+1})=\boldsymbol{\sigma}_{i}(t_{n})+\Delta\beta_{1i}\boldsymbol{e}_{1i}\times\boldsymbol{\sigma}_{i}(t_{n}) + \Delta\beta_{2i}\boldsymbol{e}_{2i}\times\boldsymbol{\sigma}_{i}(t_{n}).
\end{eqnarray}

Replacing $\boldsymbol{\sigma}_{i}(t_{n})$ by $\frac{1}{2}(\boldsymbol{\sigma}_{i}(t_{n})+\boldsymbol{\sigma}_{i}(t_{n+1}))$ on the right hand side leads to
\begin{eqnarray}
\boldsymbol{\sigma}_{i}(t_{n+1})&=&\boldsymbol{\sigma}_{i}(t_{n})+\Delta\beta_{1i}\boldsymbol{e}_{1i} \!\times \! \frac{1}{2}[\boldsymbol{\sigma}_{i}(t_{n})+\boldsymbol{\sigma}_{i}(t_{n+1})]\nonumber
\\
&&+ \Delta\beta_{2i}\boldsymbol{e}_{2i}\times\frac{1}{2}[\boldsymbol{\sigma}_{i}(t_{n})+\boldsymbol{\sigma}_{i}(t_{n+1})],
\end{eqnarray}
where it is straightforward to see that the vectors $\boldsymbol{\sigma}_{i}(t_{n+1})-\boldsymbol{\sigma}_{i}(t_{n})$ and $\boldsymbol{\sigma}_{i}(t_{n+1})+\boldsymbol{\sigma}_{i}(t_{n})$ are orthogonal, therefore the magnitude of the spin remains constant after the time step: $\boldsymbol{\sigma}_{i}^{2}(t_{n+1})=\boldsymbol{\sigma}_{i}^{2}(t_{n})$. This method is called semi-implicit in \cite{Mentink} because in order to calculate the value of $\boldsymbol{\sigma}_{i}(t_{n+1})$, a linear equation has to be solved; however, the solution of this equation is remarkably simpler than in the case where $\Delta\beta_{1i}$ and $\Delta\beta_{2i}$ also depend on $\boldsymbol{\sigma}_{i}(t_{n+1})$, which would be the truly implicit scheme.

The semi-implicit method proposed by Mentink \etal\cite{Mentink} can be rewritten in the local coordinate system with the positive $z$ axis pointing along $\boldsymbol{\sigma}_{i}(t_{n})$. This method is referenced as the two-step scheme in the paper. It has the form
\begin{eqnarray}
\Delta\tilde{\beta}_{2i}&=&\frac{\gamma'}{M_{i}}\frac{\partial E}{\partial \beta_{1i}}\Delta t- \alpha\frac{\gamma'}{M_{i}}\frac{\partial E}{\partial \beta_{2i}}\Delta t\nonumber
\\
&&+ \gamma' \sqrt{2D_{i}}\boldsymbol{e}_{2i}\Delta\boldsymbol{W}_{i} + \alpha\gamma' \sqrt{2D_{i}}\boldsymbol{e}_{1i}\Delta\boldsymbol{W}_{i},\label{twostep1}
\\
\Delta\tilde{\beta}_{1i}&=&-\frac{\gamma'}{M_{i}}\frac{\partial E}{\partial \beta_{2i}}\Delta t- \alpha\frac{\gamma'}{M_{i}}\frac{\partial E}{\partial \beta_{1i}}\Delta t\nonumber
\\
&&+ \gamma' \sqrt{2D_{i}}\boldsymbol{e}_{1i}\Delta\boldsymbol{W}_{i} - \alpha\gamma' \sqrt{2D_{i}}\boldsymbol{e}_{2i}\Delta\boldsymbol{W}_{i},
\end{eqnarray}
\begin{eqnarray}
\boldsymbol{\tilde{\sigma}}_{i}(t_{n})&=&\Bigg\{\Big[1-\frac{1}{4}\big(\frac{\Delta\tilde{\beta}_{1i}}{2}\big)^{2}-\frac{1}{4}\big(\frac{\Delta\tilde{\beta}_{2i}}{2}\big)^{2}\Big]\boldsymbol{\sigma}_{i}(t_{n})\nonumber
\\
&&+\frac{1}{2}\Delta\tilde{\beta}_{2i}\boldsymbol{e}_{1i}-\frac{1}{2}\Delta\tilde{\beta}_{1i}\boldsymbol{e}_{2i}\Bigg\}\nonumber
\\
&&\times\Big[1+\frac{1}{4}\big(\frac{\Delta\tilde{\beta}_{1i}}{2}\big)^{2}+\frac{1}{4}\big(\frac{\Delta\tilde{\beta}_{2i}}{2}\big)^{2}\Big]^{-1},
\end{eqnarray}
\begin{eqnarray}
\boldsymbol{\tilde{e}}_{2i}&=&\Bigg\{\Big[1-\frac{1}{4}\big(\frac{\Delta\tilde{\beta}_{1i}}{2}\big)^{2}\Big]\boldsymbol{e}_{2i}+\frac{1}{2}\Delta\tilde{\beta}_{1i}\boldsymbol{\sigma}_{i}\Bigg\}\Big[1+\frac{1}{4}\big(\frac{\Delta\tilde{\beta}_{1i}}{2}\big)^{2}\Big]^{-1},
\\
\boldsymbol{\tilde{e}}_{1i}&=&\Bigg\{\Big[1-\frac{1}{4}\big(\frac{\Delta\tilde{\beta}_{2i}}{2}\big)^{2}\Big]\boldsymbol{e}_{1i}-\frac{1}{2}\Delta\tilde{\beta}_{2i}\boldsymbol{\sigma}_{i}\Bigg\}\Big[1+\frac{1}{4}\big(\frac{\Delta\tilde{\beta}_{2i}}{2}\big)^{2}\Big]^{-1},
\end{eqnarray}
\begin{eqnarray}
\Delta\beta_{2i}&=&\frac{\gamma'}{M_{i}}\frac{\partial E}{\partial \tilde{\beta}_{1i}}\Delta t- \alpha\frac{\gamma'}{M_{i}}\frac{\partial E}{\partial \tilde{\beta}_{2i}}\Delta t\nonumber
\\
&&+ \gamma' \sqrt{2D_{i}}\boldsymbol{\tilde{e}}_{2i}\Delta\boldsymbol{W}_{i} + \alpha\gamma' \sqrt{2D_{i}}\boldsymbol{\tilde{e}}_{1i}\Delta\boldsymbol{W}_{i},
\\
\Delta\beta_{1i}&=&-\frac{\gamma'}{M_{i}}\frac{\partial E}{\partial \tilde{\beta}_{2i}}\Delta t- \alpha\frac{\gamma'}{M_{i}}\frac{\partial E}{\partial \tilde{\beta}_{1i}}\Delta t\nonumber
\\
&&+ \gamma' \sqrt{2D_{i}}\boldsymbol{\tilde{e}}_{1i}\Delta\boldsymbol{W}_{i} - \alpha\gamma' \sqrt{2D_{i}}\boldsymbol{\tilde{e}}_{2i}\Delta\boldsymbol{W}_{i},
\end{eqnarray}
\begin{eqnarray}
\boldsymbol{B}_{i}=-\frac{1}{2}\big(\Delta\beta_{1i}\boldsymbol{\tilde{e}}_{1i}+\Delta\beta_{2i}\boldsymbol{\tilde{e}}_{2i}\big),
\\
\boldsymbol{A}_{i}=\boldsymbol{\sigma}_{i}(t_{n})+\boldsymbol{\sigma}_{i}(t_{n})\times\boldsymbol{B}_{i},
\\
\boldsymbol{\sigma}_{i}(t_{n+1}) = \Big[\boldsymbol{A}_{i}+\boldsymbol{A}_{i}\times\boldsymbol{B}_{i}+(\boldsymbol{A}_{i}\boldsymbol{B}_{i})\boldsymbol{B}_{i}\Big] \Big(1+\boldsymbol{B}_{i}^{2}\Big)^{-1},\label{twostep2}
\end{eqnarray}
where we explicitly provided the solutions of the linear equations needed in the semi-implicit calculation.

Similar to the Heun scheme\cite{Palacios}, the above procedure is a predictor-corrector method; however, the predictor scheme gives a first approximation to $\boldsymbol{\sigma}_{i}\big(\frac{1}{2}(t_{n}+t_{n+1})\big)$ instead of $\boldsymbol{\sigma}_{i}(t_{n+1})$, therefore in the first step only a smaller rotation happens with the angles $\frac{1}{2}\Delta\tilde{\beta}_{1i},\frac{1}{2}\Delta\tilde{\beta}_{2i}$. The random variables $\Delta W_{i}^{r}$, where $r$ denotes Descartes components, are calculated from independent, identically distributed standard normal random variables $\xi_{i}^{r}$ as $\Delta W_{i}^{r}=\sqrt{\Delta t}\xi_{i}^{r}$, where $\Delta t=t_{n+1}-t_{n}$ is the time step, being fixed during the simulation. This method converges weakly to the solution of the equation with order $\delta=1$, just like the Heun method. However, it was demonstrated in \cite{Mentink} that it remains more stable than the Heun method when increasing the time step.

To present the one-step scheme we introduce the shorthand notations
\begin{eqnarray}
x_{2i}=\frac{\gamma'}{M_{i}}\frac{\partial E}{\partial \beta_{1i}} - \alpha \frac{\gamma'}{M_{i}}\frac{\partial E}{\partial \beta_{2i}},
\\
x_{1i}=-\frac{\gamma'}{M_{i}}\frac{\partial E}{\partial \beta_{2i}} - \alpha \frac{\gamma'}{M_{i}}\frac{\partial E}{\partial \beta_{1i}},
\\
x_{2j2i}=\frac{\gamma'}{M_{i}}\frac{\partial^{2} E}{\partial \beta_{2j} \partial \beta_{1i}} - \alpha \frac{\gamma'}{M_{i}}\frac{\partial^{2} E}{\partial \beta_{2j} \partial \beta_{2i}},
\\
x_{1j2i}=\frac{\gamma'}{M_{i}}\frac{\partial^{2} E}{\partial \beta_{1j}\partial \beta_{1i}} - \alpha \frac{\gamma'}{M_{i}}\frac{\partial^{2} E}{\partial \beta_{1j} \partial \beta_{2i}},
\\
x_{2j1i}=-\frac{\gamma'}{M_{i}}\frac{\partial^{2} E}{\partial \beta_{2j}\partial \beta_{2i}} - \alpha \frac{\gamma'}{M_{i}}\frac{\partial^{2} E}{\partial \beta_{2j} \partial \beta_{1i}},
\\
x_{1j1i}=-\frac{\gamma'}{M_{i}}\frac{\partial^{2} E}{\partial \beta_{1j} \partial \beta_{2i}} - \alpha \frac{\gamma'}{M_{i}}\frac{\partial^{2} E}{\partial \beta_{1j}\partial \beta_{1i}},
\end{eqnarray}
\begin{eqnarray}
s_{2i}^{r}=\gamma'\sqrt{2D_{i}}(e_{2i}^{r}+\alpha e_{1i}^{r}),
\\
s_{1i}^{r}=\gamma'\sqrt{2D_{i}}(e_{1i}^{r}-\alpha e_{2i}^{r}),
\\
s_{2i2i}^{r}=s_{1i1i}^{r}=-\alpha\gamma'\sqrt{2D_{i}}\sigma_{i}^{r},
\\
s_{1i2i}^{r}=-s_{2i1i}^{r}=\gamma'\sqrt{2D_{i}}\sigma_{i}^{r},
\end{eqnarray}
and the approximate Stratonovich integrals
\begin{eqnarray}
\hat{J}_{(0)}=\Delta t,
\\
\hat{J}_{(ir)}=\Delta W_{i}^{r}=\sqrt{\Delta t}\xi_{1i}^{r},
\\
\hat{J}_{(0,0)}=\frac{\Delta t^{2}}{2},
\\
\hat{J}_{(ir,ir)}=\frac{(\Delta W_{i}^{r})^{2}}{2},
\end{eqnarray}
\begin{eqnarray}
\hat{J}_{(0,ir)}=\frac{1}{2}\Delta t^{\frac{3}{2}}(\xi_{1i}^{r}-\frac{1}{\sqrt{3}}\xi_{2i}^{r}),
\\
\hat{J}_{(ir,0)}=\frac{1}{2}\Delta t^{\frac{3}{2}}(\xi_{1i}^{r}+\frac{1}{\sqrt{3}}\xi_{2i}^{r}),
\\
\hat{J}_{(ir,ir')}=\frac{1}{2} \Delta t(\xi_{1i}^{r}\xi_{1i}^{r'}+\xi_{3i}^{r}\xi_{3i}^{r'})  \qquad {\rm if} \: r>r',
\\
\hat{J}_{(ir',ir)}=\frac{1}{2} \Delta t(\xi_{1i}^{r}\xi_{1i}^{r'}-\xi_{3i}^{r}\xi_{3i}^{r'}) \qquad {\rm if} \: r>r',
\end{eqnarray}
where the $\xi_{1i}^{r}, \xi_{2i}^{r}$ and $\xi_{3i}^{r}$ random variables are standard normally distributed and independent for different indices $1, 2, 3$,  lattice points $i$, Descartes components $r$ and time steps. For comparison, in the two-step scheme only the Stratonovich integrals $\hat{J}_{(0)}=\Delta t$ and $\hat{J}_{(ir)}=\Delta W_{i}^{r}=\sqrt{\Delta t}\xi_{i}^{r}$ have to be calculated.

With the above notations, the one-step numerical scheme used by us to solve equations (\ref{LLGloc1})-(\ref{LLGloc2}) has the form
\begin{eqnarray}
\Delta \beta_{2i}&=& \: x_{2i}\hat{J}_{(0)} + \sum_{r}s_{2i}^{r}\hat{J}_{(i r)} + \sum_{j}(x_{2j}x_{2j2i}+x_{1j}x_{1j2i})\hat{J}_{(0,0)} \nonumber
\\
&&+ \sum_{j,r}(s_{2j}^{r}x_{2j2i}+s_{1j}^{r}x_{1j2i})\hat{J}_{(jr,0)} + \sum_{r}(x_{2i}s_{2i2i}^{r}+x_{1i}s_{1i2i}^{r})\hat{J}_{(0,ir)} \nonumber
\\
&&+ \sum_{r,r'}(s_{2i2i}^{r}s_{2i}^{r'}+s_{1i2i}^{r}s_{1i}^{r'})\hat{J}_{(ir',ir)},
\end{eqnarray}
\begin{eqnarray}
\Delta \beta_{1i}&=& \: x_{1i}\hat{J}_{(0)} + \sum_{r}s_{1i}^{r}\hat{J}_{(ir)} +\sum_{j} (x_{2j}x_{2j1i}+x_{1j}x_{1j1i})\hat{J}_{(0,0)} \nonumber
\\
&&+ \sum_{j,r}(s_{2j}^{r}x_{2j1i}+s_{1j}^{r}x_{1j1i})\hat{J}_{(jr,0)} + \sum_{r}(x_{2i}s_{2i1i}^{r}+x_{1i}s_{1i1i}^{r})\hat{J}_{(0,ir)} \nonumber
\\
&&+ \sum_{r,r'}(s_{2i1i}^{r}s_{2i}^{r'}+s_{1i1i}^{r}s_{1i}^{r'})\hat{J}_{(ir',ir)},\label{onestep1}
\end{eqnarray}
\begin{eqnarray}
\boldsymbol{\sigma}_{i}(t_{n+1})&=&\Bigg\{\Big(1-\frac{1}{4}\Delta\beta_{1i}^{2}-\frac{1}{4}\Delta\beta_{2i}^{2}\Big)\boldsymbol{\sigma}_{i}(t_{n})+\Delta\beta_{2i}\boldsymbol{e}_{1i}-\Delta\beta_{1i}\boldsymbol{e}_{2i}\Bigg\}\nonumber
\\
&&\times\Big[1+\frac{1}{4}\Delta\beta_{1i}^{2}+\frac{1}{4}\Delta\beta_{2i}^{2}\Big]^{-1}.\label{onestep2}
\end{eqnarray}

When calculating the values of the spins at the next time step, the same algorithm was used with the vector products as before, thereby conserving the length of the spins. The second derivatives of the energy functional ($x_{2j2i}, x_{2j1i}, x_{1j2i}, x_{1j1i}$) were taken from (\ref{secder}). As noted in section \ref{numint}, the calculation of these quantities from first principles takes less time since the scattering path operator needed for the first and second derivatives of the energy must be determined for only one magnetic configuration. In the deterministic limit, that is at $T=0$, this method is a second-order scheme, just like the deterministic Heun scheme or the semi-implicit two-step scheme. At finite temperatures the one-step scheme also has weak order of convergence $\delta=1$.

The simplified one-step scheme has the form
\begin{eqnarray}
\Delta\beta_{2i}&=&\frac{\gamma'}{M_{i}}\frac{\partial E}{\partial \beta_{1i}}\Delta t- \alpha\frac{\gamma'}{M_{i}}\frac{\partial E}{\partial \beta_{2i}}\Delta t\nonumber
\\
&&+ \gamma' \sqrt{2D_{i}}\boldsymbol{e}_{2i}\Delta\boldsymbol{W}_{i} + \alpha\gamma' \sqrt{2D_{i}}\boldsymbol{e}_{1i}\Delta\boldsymbol{W}_{i},\label{simp1}
\\
\Delta\beta_{1i}&=&-\frac{\gamma'}{M_{i}}\frac{\partial E}{\partial \beta_{2i}}\Delta t- \alpha\frac{\gamma'}{M_{i}}\frac{\partial E}{\partial \beta_{1i}}\Delta t\nonumber
\\
&&+ \gamma' \sqrt{2D_{i}}\boldsymbol{e}_{1i}\Delta\boldsymbol{W}_{i} - \alpha\gamma' \sqrt{2D_{i}}\boldsymbol{e}_{2i}\Delta\boldsymbol{W}_{i},
\end{eqnarray}
\begin{eqnarray}
\boldsymbol{\sigma}_{i}(t_{n+1})&=&\Bigg\{\Big(1-\frac{1}{4}\Delta\beta_{1i}^{2}-\frac{1}{4}\Delta\beta_{2i}^{2}\Big)\boldsymbol{\sigma}_{i}(t_{n})+\Delta\beta_{2i}\boldsymbol{e}_{1i}-\Delta\beta_{1i}\boldsymbol{e}_{2i}\Bigg\}\nonumber
\\
&&\times\Big[1+\frac{1}{4}\Delta\beta_{1i}^{2}+\frac{1}{4}\Delta\beta_{2i}^{2}\Big]^{-1},\label{simp2}
\end{eqnarray}
where $\Delta W_{i}^{r}=\sqrt{\Delta t}\, \xi_{i}^{r}$, with the same quantities as in the two-step scheme. Importantly, equation (\ref{simp2}) conserves the magnitude of the spin vectors.
A simple Euler method using the coefficients from the Stratonovich form of the equation is not convergent at all\cite{Palacios,Kloeden}, but this modification corrigates the error and it also has weak order of convergence $\delta=1$. On the other hand, the earlier two methods are in a certain sense much \textquotedblleft closer\textquotedblright\, to a second-order scheme than the one based on the Euler method, since the deterministic limit of those methods has second order of convergence, while the deterministic Euler method is only of first order. Probably this is the reason why the simplified scheme requires a  $100$ times smaller time step than the other two schemes as shown in section \ref{model}.

\subsection{The model Hamiltonian\label{appendix2}}
Considering the simple spin model
\begin{eqnarray}
E=J\sum_{i=1}^{N-1}\boldsymbol{\sigma}_{i}\boldsymbol{\sigma}_{i+1}+K\sum_{i=1}^{N}(\sigma_{i}^{x})^{2},
\end{eqnarray}
the first and second derivatives of the energy in the local coordinate system can be given as
\begin{eqnarray}
\frac{\partial E}{\partial \beta_{2i}}&=&J\sum_{j=i \pm 1}\boldsymbol{e}_{1i}\boldsymbol{\sigma}_{j}+2Ke_{1i}^{x}\sigma_{i}^{x},
\\
\frac{\partial E}{\partial \beta_{1i}}&=&-J\sum_{j=i \pm 1}\boldsymbol{e}_{2i}\boldsymbol{\sigma}_{j}-2Ke_{2i}^{x}\sigma_{i}^{x},
\end{eqnarray}
and
\begin{eqnarray}
\frac{\partial ^{2} E}{\partial \beta_{2j} \partial \beta_{2i}}=J\boldsymbol{e}_{1i}\boldsymbol{e}_{1j} &&\qquad {\rm if}\, j=i \pm 1,\label{Jcalc1}
\\
\frac{\partial ^{2} E}{\partial \beta_{2j} \partial \beta_{1i}}=-J\boldsymbol{e}_{2i}\boldsymbol{e}_{1j} &&\qquad {\rm if}\, j=i \pm 1,
\\
\frac{\partial ^{2} E}{\partial \beta_{1j} \partial \beta_{2i}}=-J\boldsymbol{e}_{1i}\boldsymbol{e}_{2j} &&\qquad {\rm if}\, j=i \pm 1,
\\
\frac{\partial ^{2} E}{\partial \beta_{1j} \partial \beta_{1i}}=J\boldsymbol{e}_{2i}\boldsymbol{e}_{2j} &&\qquad {\rm if}\, j=i \pm 1,\label{Jcalc2}
\end{eqnarray}
\begin{eqnarray}
\frac{\partial ^{2} E}{\partial \beta_{2i}^{2}}=-J\sum_{j=i \pm 1}\boldsymbol{\sigma}_{i}\boldsymbol{\sigma}_{j}-2K(\sigma_{i}^{x})^{2}+2K(e_{1i}^{x})^{2},
\\
\frac{\partial ^{2} E}{\partial \beta_{1i}^{2}}=-J\sum_{j=i \pm 1}\boldsymbol{\sigma}_{i}\boldsymbol{\sigma}_{j}-2K(\sigma_{i}^{x})^{2}+2K(e_{2i}^{x})^{2},
\\
\frac{\partial ^{2} E}{\partial \beta_{1i} \partial \beta_{2i}}=\frac{\partial ^{2} E}{\partial \beta_{2i} \partial \beta_{1i}}=-2Ke_{1i}^{x}e_{2i}^{x}. \label{ed2}
\end{eqnarray}

The above quantities are necessary in the model calculations testing the stability of the numerical integration schemes in section \ref{model} and in calculating the flipping times in section \ref{Co10Au001}. Moreover, if the second derivatives of the energy are calculated from the ab initio method, see (\ref{secder}), expressions (\ref{Jcalc1})--(\ref{ed2})  provide possible alternatives to determine the exchange coefficient $J$ and the anisotropy constant $K$ for a suitable model Hamiltonian. Clearly, this procedure is ambiguous, therefore in section \ref{Co10Au001} we took an average
of the $J$ and $K$ values obtained from different types of second derivatives.

\subsection{Approximating the error of the simulations \label{confidence}}

Let $X$ be a physical quantity that has to be determined from the simulations. After running the simulations $N$ times and taking the values of $X$ at the end ($X_{i}, i=1,\dots, N$) the average value
\begin{eqnarray}
X_{av}=\frac{1}{N}\sum_{i}X_{i}
\end{eqnarray}
as well as the empirical variance
\begin{eqnarray}
{\rm Var}(X)=\frac{1}{N-1}\sum_{i}\big(X_{i}-X_{av}\big)^{2}
\end{eqnarray}
are calculated. If $N$ is large enough, it can be assumed that $X_{av}$ is of Gaussian distribution with variance $\frac{1}{N}{\rm Var}(X)$. Therefore the expectation value $\langle X \rangle$ falls into a confidence interval
around $X_{av}$,
\begin{eqnarray}
\langle X \rangle &\in& \Bigg(X_{av}-1.96\sqrt{\frac{1}{N}{\rm Var}(X)},X_{av}+1.96\sqrt{\frac{1}{N}{\rm Var}(X)}\Bigg)
\end{eqnarray}
with probability $0.95$.

\subsection{Determining the normal modes of the system\label{normalmodes}}

Here we give a general scheme to find the normal modes of an arbitrary spin system described by the Landau-Lifshitz-Gilbert equations. Equations (\ref{LLGloc1})-(\ref{LLGloc2}) without thermal noise and damping have the form
\begin{eqnarray}
M_{i}\frac{\partial \beta_{2i}}{\partial t}&=&\gamma\frac{\partial E}{\partial \beta_{1i}}, \label{LLGsimp1}
\\
M_{i}\frac{\partial \beta_{1i}}{\partial t}&=&-\gamma\frac{\partial E}{\partial \beta_{2i}}, \label{LLGsimp2}
\end{eqnarray}
which is analogous to the canonical equations in Hamiltonian mechanics. Introducing $p_{i}=\sqrt{\frac{M_{i}}{\gamma}}\beta_{1i}$ standing for a generalized momentum and $q_{i}=\sqrt{\frac{M_{i}}{\gamma}}\beta_{2i}$ for the corresponding generalized coordinate, the energy can be expanded up to second order terms close to the ground state in these generalized coordinates and momenta as
\begin{eqnarray}
E=E_{0} + \frac{1}{2}\sum_{i,j}\Big(A_{ij}p_{i}p_{j}+B_{ij}p_{i}q_{j}+B_{ji}p_{j}q_{i}+C_{ij}q_{i}q_{j}\Big),\label{energyeq}
\end{eqnarray}
where
\begin{eqnarray}
A_{ij}&=&\frac{\partial ^2 E}{\partial p_{i} \partial p_{j}}=\frac{\gamma}{\sqrt{M_{i}M_{j}}}\frac{\partial ^2 E}{\partial \beta_{1i} \partial \beta_{1j}},
\\
B_{ij}&=&\frac{\partial ^2 E}{\partial p_{i} \partial q_{j}}=\frac{\gamma}{\sqrt{M_{i}M_{j}}}\frac{\partial ^2 E}{\partial \beta_{1i} \partial \beta_{2j}},
\\
C_{ij}&=&\frac{\partial ^2 E}{\partial q_{i} \partial q_{j}}=\frac{\gamma}{\sqrt{M_{i}M_{j}}}\frac{\partial ^2 E}{\partial \beta_{2i} \partial \beta_{2j}} \: .
\end{eqnarray}

The equations of motion can then be derived from equations (\ref{LLGsimp1}) and (\ref{LLGsimp2}),
\begin{eqnarray}
\dot{q}_{i}&=&\sum_{j}\Big(A_{ij}p_{j}+B_{ij}q_{j}\Big), \nonumber
\\
\dot{p}_{i}&=&-\sum_{j}\Big(B_{ji}p_{j}+C_{ij}q_{j}\Big).\label{motioneq}
\end{eqnarray}

Equations (\ref{energyeq}) and (\ref{motioneq}) can be rewritten using matrix notation,
$(\boldsymbol{p}, \boldsymbol{q}) = \left( \{ p_i \}, \{ q_i \}  \right)$, as
\begin{eqnarray}
E=E_{0}+\frac{1}{2}\left[\begin{array}{cc}\boldsymbol{q}^{T} & \boldsymbol{p}^{T}\end{array}\right]\left[\begin{array}{cc}\boldsymbol{C} & \boldsymbol{B}^{T} \\ \boldsymbol{B} & \boldsymbol{A}\end{array}\right]\left[\begin{array}{c}\boldsymbol{q} \\ \boldsymbol{p}\end{array}\right],\label{energyeq2}
\\
\left[\begin{array}{c}\dot{\boldsymbol{q}} \\ \dot{\boldsymbol{p}}\end{array}\right]=\left[\begin{array}{cc} \boldsymbol{B} & \boldsymbol{A} \\ -\boldsymbol{C} & -\boldsymbol{B}^{T} \end{array}\right]\left[\begin{array}{c}\boldsymbol{q} \\ \boldsymbol{p}\end{array}\right] \: . \label{motioneq2}
\end{eqnarray}

Assuming the form $\boldsymbol{q}(t),\boldsymbol{p}(t) \propto \boldsymbol{q}_{k},\boldsymbol{p}_{k} \rme^{\rmi\omega_{k}t}$ for the normal modes, the equation of motion (\ref{motioneq2}) simplifies to the eigenvalue
equation,
\begin{eqnarray}
\omega_{k}\left[\begin{array}{c}\boldsymbol{q}_{k} \\ \boldsymbol{p}_{k}\end{array}\right]&=&\left[\begin{array}{cc} 0 & -\rmi \\ \rmi & 0 \end{array}\right]\left[\begin{array}{cc}\boldsymbol{C} & \boldsymbol{B}^{T} \\ \boldsymbol{B} & \boldsymbol{A}\end{array}\right]\left[\begin{array}{c}\boldsymbol{q}_{k} \\ \boldsymbol{p}_{k}\end{array}\right]\nonumber
\\
&=&\sigma_{y}\boldsymbol{H}\left[\begin{array}{c}\boldsymbol{q}_{k} \\ \boldsymbol{p}_{k}\end{array}\right].
\end{eqnarray}
where $\omega_{k}$ is the eigenvalue of the matrix $\sigma_{y}\boldsymbol{H}$, with the Pauli matrix $\sigma_{y}$ and the matrix $\boldsymbol{H}$ appearing on the right-hand side of (\ref{energyeq2}). $\boldsymbol{H}$ is a positive definite matrix if the ground state corresponds to an energy minimum, therefore $\boldsymbol{H}^{\frac{1}{2}}$ exists, it is invertible, and $\sigma_{y}\boldsymbol{H}$ has the same eigenvalues as $\boldsymbol{H}^{\frac{1}{2}}\sigma_{y}\boldsymbol{H}^{\frac{1}{2}}$. Since the latter one is a self-adjoint matrix, all the $\omega_{k}$ eigenvalues are real numbers, thus they represent the normal modes of the system. On the other hand, since the purely imaginary $\rmi\omega_{k}$ is an eigenvalue of the real valued matrix appearing in (\ref{motioneq2}), $-\rmi\omega_{k}$ must also be an eigenvalue, therefore the normal modes always appear in $\pm\omega_{k}$ pairs.

The calculation does not change considerably if the matrix $\boldsymbol{H}$ has zero eigenvalues. In this case $\sigma_{y}\boldsymbol{H}$ also has zero eigenvalues with the same eigenvectors as $\boldsymbol{H}$, and one can determine the nonzero eigenvalues on the subspace where $\boldsymbol{H}$ is strictly positive definite, using the algorithm given above.

A similar method for calculating the normal modes (magnon spectrum) of a layered system with discrete translational invariance in the plane is given in \cite{Erickson}, where the quantum mechanical equation of motion was used instead of equations (\ref{LLGsimp1}) and (\ref{LLGsimp2}).

\end{document}